\documentclass[12pt,twoside,pointlessnumbers,smallheadings]{revtex4-1}
\usepackage{amsmath}
\usepackage{amssymb}
\usepackage{color}

\usepackage{hangcaption,fancybox,feynmp}
\usepackage{indentfirst}
\usepackage{graphicx}
\usepackage{epsfig,subfigure,psfrag}
\usepackage{CJK}
\usepackage{mathrsfs}

\begin{document}


\title{ Discriminating Different $Z^\prime$s via Asymmetries at the LHC }
\author{Zhong-qiu Zhou$^{1}$\footnote{E-mail:zhongqiu\_zhou@pku.edu.cn}, Bo
Xiao$^{1}$\footnote{E-mail:homenature@pku.edu.cn},
 You-kai Wang
$^{1}$\footnote{E-mail:wangyk@pku.edu.cn}, Shou-hua
Zhu$^{1,2}$\footnote{E-mail:shzhu@pku.edu.cn}}

\affiliation{ $ ^1$ Institute of Theoretical Physics $\&$ State Key
Laboratory of Nuclear Physics and Technology, Peking University,
Beijing 100871, China \\
$ ^2$ Center for High Energy Physics, Peking University, Beijing
100871, China }

\date{\today}
\maketitle

\begin{center}

{\bf Abstract }

\begin{minipage}{16cm}
{\small  \hskip 0.25cm

In practice the asymmetry, which is defined based on the angular distribution of the final states in
scattering or decay processes,
can be utilized to scrutinize underlying dynamics in and/or beyond the standard model (BSM). As
one of the possible BSM physics which might be discovered early at the LHC, extra neutral gauge bosons $Z^\prime$s are theoretical well motivated. Once $Z^\prime$s
are discovered at the LHC, it is crucial to discriminate different $Z^\prime$s in
various BSM. In principle such task can be accomplished by measuring
the angular distribution of the final states which are produced via $Z^\prime$-mediated processes.
In the real data analysis, asymmetry is always adopted.
In literature several asymmetries have been proposed
at the LHC.  Based on these works, we
stepped further on to study how to optimize the asymmetries in the
left-right model and the sequential standard model, as the examples of BSM.
In this paper, we examined four kinds of asymmetries, namely rapidity-dependent forward-backward asymmetry,
one-side forward-backward asymmetry, central charge asymmetry and
edge charge asymmetry (see text for details), with $\ell^+\ell^-$ ($\ell=e,\mu$), $b\bar b$ and $t\bar t$
as the final states.
In the calculations with $b\bar b$ and $t\bar t$ final states, the QCD-induced
higher order contributions to the asymmetric cross section were also included.
For each kind of final states, we estimated the four kinds of asymmetries and
especially the optimal cut usually associated with the definition of the asymmetry. Our numerical results indicated that the capacity to discriminate $Z^\prime$ models can be improved by imposing the optimal
cuts.

 }

\end{minipage}
\end{center}


\newpage

\section{Introduction\label{introduction}}

LHC is a powerful machine for discovering new particle and examining its
couplings if the new particle is at $\mathscr{O}$(TeV) or below. In the physics
beyond the standard model (BSM), there are usually new particles. For example in the
simplest case, besides the standard model (SM) gauge group $SU(3)_C \otimes SU(2)_L \otimes
U(1)_Y$, there can be an extra abelian gauge group $U(1)$ which implies the existence of
the extra gauge boson dubbed as $Z^\prime$. If the mass of $Z^\prime$ is not so heavy,
it can be discovered early at the LHC. Similar to the case of $J/\psi$ discovery, $Z^\prime$ might
show up as the di-muon resonance. In fact numerous phenomenological studies on $Z^\prime$
at the LHC have been  carried out.
After the discovery it is very important to
study its spin, couplings etc. in order to fix the nature of the physics behind the new particle.
In principle such detail information can be obtained
via the precise measurement for the angular distributions of the final states into which
$Z^\prime$s decay. However in practice the asymmetry is usually utilized to investigate
the detail properties of the new particle, provided that the data sample is usually limited
in the real experiments. The measurement of the asymmetry at the LEP (Tevatron), as the charge asymmetric electron-positron (proton-antiproton) collider, has shed light on the knowledge of the SM and constrained the BSM severely. The LHC, as the charge symmetric proton-proton collider,
has unique feature to define and measure the asymmetry.

In literature, several asymmetry definitions have already been
proposed to study the underlying dynamics of the BSM. Note that asymmetry is applicable
to all kinds of new physics, not limited to $Z^\prime$. In this paper $Z^\prime$ is
only taken as an example.
In order to distinguish different $Z^\prime$ models, forward-backward
asymmetry  at the hadron
collider is one of the most important tools which was suggested in
1984 in the study of $Z^\prime$ physics \cite{PhysRevD.30.1470}. Afterwards, it was widely used in
the $Z^\prime$ study \cite{hepph.9504216, PhysRevD.55.161,
PhysLettB.583.111, PhysRevLett.101.151803, PhysRevD.77.115004,
RevModPhys.81.1199, arXiv.0910.1334, PhysRevD.80.075014,
arXiv.1006.2845, PhysRevD.36.979, NuclPhysB.287.419,
PhysRevD.35.2893, PhysLettB.180.163, PhysRevD.37.643,
PhysLettB.221.85, PhysLettB.202.411, PhysRevD.46.290,
PhysRevD.46.14, PhysRevD.45.278, Feruglio:1991mm,
Verzegnassi:1987zz, Rosner:1986ad, PhysRevD.54.1078,
PhysRevD.48.R969, Abdallah:2005ph, arXiv.0911.4294,
PhysRevD.82.115020} and has been developed into more convenient
forms. In fact at the LHC, there are several other types of asymmetry definitions
 \cite{PhysRevLett.81.49,
PhysRevD.59.054017, PhysRevD.77.014003, PhysRevD.78.094018,
PhysRevD.82.094011, arXiv.1011.1428, arXiv.1101.2507}. It is crucial
to compare them and find out the most suitable one for the
specific purpose, eg. to study the specific couplings between ${Z^\prime}$ and the SM fermions. Furthermore, each
type of asymmetry definition contains characteristic cuts
which should be chosen properly to make the asymmetry most
significant. The most suitable type of
asymmetry and the most proper cuts associated with asymmetry are different for different
physics. In this paper, we are going to investigate the optimal cuts  for identifying the
$Z^\prime$ in different models at the LHC. In this paper two kinds of $Z^\prime$ test models namely the left-right (LR) and sequential standard model (SSM) are adopted.

In the previous study \cite{arXiv.0910.1334}, the authors utilized the forward-backward asymmetry
defined by themselves in order to identify the
different $Z^\prime$s, which can decay into $\mu^+\mu^-(e^+e^-)$,
$b\bar b$ as well as  $t\bar t$. For the final states of $\mu^+\mu^-(e^+e^-)$,
asymmetry is calculated for the
on-peak data sample namely the invariant mass of the charged lepton pair lying within $M_{\rm{{Z^\prime}}}-3\Gamma$ and $M_{\rm{{Z^\prime}}}+3\Gamma$, as well as
the off-peak data sample with invariant mass lying within $2/3M_{\rm{{Z^\prime}}}$ and $M_{\rm{{Z^\prime}}}-3\Gamma$.
Here $\Gamma$ is the total width of $Z^\prime$.
 For the
$b\bar b$ and $t\bar t$ final states, only asymmetry of
on-peak data sample with quark pair invariant mass within $M_{\rm{{Z^\prime}}}-2.5\Gamma$ and $M_{\rm{{Z^\prime}}}+2.5\Gamma$ was
calculated. In this paper we extended the above analysis to more asymmetries defined recently, namely
one-side forward-backward asymmetry, central charge asymmetry and edge charge asymmetry. In our calculation for
the quarks as the final states, we included also the contributions from the higher-order QCD-induced effects.
Moreover we investigated the optimal conditions which make the asymmetry more significant.
Explicitly, in our paper we are going to scrutinize which
type of asymmetry and the corresponding cuts are the most
suitable ones for the $\mu^+\mu^-(e^+e^-)$ on/off-peak events
the $b\bar b$ and $t\bar t$ events respectively.

In section \ref{LHC FBA}, different asymmetries at the LHC are briefly
described. In section \ref{OPTIMIZATION}, we firstly calculated
four types of asymmetries at the LHC with their characteristic cuts for
the $\mu^+\mu^-(e^+e^-)$ on/off-peak events, $b\bar b$,
$t\bar t$ on-peak events respectively. Secondly, based on the calculations we optimize the asymmetry for
each case by choosing different cuts. Thirdly, we showed how to discriminate different $Z^\prime$s in LR and
SSM models utilizing the
optimized asymmetry. In section \ref{CONCLUSION}, we gave our discussions
and conclusions.

\section{Asymmetry at the LHC\label{LHC FBA}}

LHC is the {\em symmetric} proton-proton collider, thus the usual defined asymmetry is absolute zero after
integrating all kinematical region. However if one selects events in a certain kinematical region, the
asymmetry arising at partonic level will be kept.

\subsection{Forward-backward asymmetry for the partonic level process $q\bar q\rightarrow f\bar
f$}

In order to illustrate the $Z^\prime$ contribution to
asymmetry, we describe firstly the forward-backward (FB) asymmetry in the SM.
At the tree level in the SM, the FB asymmetry of $q\bar
q\rightarrow Z/\gamma^*\rightarrow l\bar l$ gets contributions from
the self conjugation of the Z-induced s-channel feynman diagram
\begin{equation}
\begin{split}
\frac{d\sigma}{d\cos\theta}\propto
\frac{g^4s^2}{c_W^4[(s-M_Z^2)^2+\Gamma_Z^2M_Z^2]}[&((g_L^l)^2+(g_R^l)^2)((g_L^q)^2+(g_R^q)^2)(1+\cos^2\theta) \\
&+2((g_L^l)^2-(g_R^l)^2)((g_L^q)^2-(g_R^q)^2)\cos\theta],
 \end{split}
 \end{equation}
 and the interference between this diagram and the
$\gamma$-induced s-channel feynman diagram
\begin{equation}
\begin{split}
\frac{d\sigma}{d\cos\theta}\propto
\frac{g^2ee_q(M_Z^2-s)s}{c_W^2[(s-M_Z^2)^2+\Gamma_Z^2M_Z^2]}[&2(g_L^l+g_R^l)(g_L^q+g_R^q)(1+\cos^2\theta) \\
&+4(g_L^l-g_R^l)(g_L^q-g_R^q)\cos\theta].
 \end{split}
 \end{equation}
Around the Z-pole, the FB asymmetry is almost determined by the first
contribution, while off the Z-pole, the second contribution plays an
important role.

After introducing $Z^\prime$, the situation becomes a little bit complicated.
On the $Z^\prime$ pole, if the coupling of the $Z^\prime$ to fermions is
\emph{not} pure vector-like nor pure axial-vector-like namely
$|g_L^f|=|g_R^f|$, there will be non-zero
contribution to FB asymmetry from the self-conjugation of the $Z^\prime$
induced s-channel feynman diagram.
Otherwise contribution from self-conjugation of $Z^\prime$
to F-B asymmetry will be zero. However for the data sample off the $Z^\prime$ pole,
FB asymmetry will be non-zero due to the interference of the
$Z^\prime$ diagram and the SM Z/$\gamma^*$
induced s-channel feynman diagrams.

From the above description, we can see clearly that the FB asymmetry
relates tightly to the chiral properties of the couplings and how to
select data samples. In the BSM which contains the $Z^\prime$, the coupling of
$Z^\prime$ and SM fermions is usually different. How
to extract the corresponding couplings via asymmetry measurement and subsequently discriminate
different BSM is the
key motivation for both the theoretical and experimental studies.

The above formulas are applicable also to $q\bar q\rightarrow b\bar
b(t\bar t)$, however there are additional important contributions to the
F-B asymmetry from the QCD high-order processes
\cite{PhysRevLett.81.49, PhysRevD.59.054017}. If one selects
the events around the $Z^\prime$ pole, the QCD high-order
contributions are suppressed. However such effect will be important
for the asymmetry of off-pole events. In this paper, such QCD-induced contributions will
be included in the analysis.

\subsection{Four asymmetries defined at the LHC}

The proton-proton collider LHC is forward-backward charge symmetric,
so the asymmetry of the fermion pairs produced at the LHC is null
if integrating over the full phase space. However, by imposing some kinematical cuts, the asymmetry
generated at the partonic level $q\bar q\rightarrow f\bar
f$ can be kept. Different types of asymmetries have been
defined. The above-mentioned FB
asymmetry \cite{PhysRevD.30.1470,hepph.9504216, PhysRevD.55.161,
PhysLettB.583.111, PhysRevLett.101.151803, PhysRevD.77.115004,
RevModPhys.81.1199, arXiv.0910.1334, PhysRevD.80.075014,
arXiv.1006.2845} which have been frequently used in $Z^\prime$ studies
contains a characteristic fermion pair rapidity $Y_{{f\bar f}}$ cut.
We refer it as rapidity dependent forward-backward asymmetry
($\rm{A_{RFB}}$) throughout this paper. The other three asymmetries which will be
investigated in this paper are
one-side forward-backward asymmetry ($\rm{A_{OFB}}$)
\cite{PhysRevD.82.094011, arXiv.1011.1428}, central charge
asymmetry ($\rm{A_C}$) \cite{PhysRevLett.81.49, PhysRevD.59.054017,
PhysRevD.77.014003, PhysRevD.78.094018}, and edge charge
asymmetry ($\rm{A_E}$) \cite{arXiv.1101.2507}. We collect their
definitions as below
\begin{equation}
A_{\rm{RFB}}(Y_{{f\bar f}}^{\rm{cut}})=\left. \frac{\sigma(|Y_{
f}|>|Y_{{\bar f}}|)-\sigma(|Y_{ f}|<|Y_{{\bar f}}|)}{\sigma(|Y_{
f}|>|Y_{{\bar f}}|)+\sigma(|Y_{f}|<|Y_{{\bar
f}}|)}\right|_{|Y_{{f\bar f}}|>Y_{{f\bar f}}^{{cut}}},
 \end{equation}
\begin{equation}
A_{\rm{OFB}}(p^{\rm{cut}}_{Z,{f\bar f}})=\left. \frac{\sigma(|Y_{
f}|>|Y_{{\bar f}}|)-\sigma(|Y_{f}|<|Y_{{\bar f}}|)}{\sigma(|Y_{
f}|>|Y_{{\bar f}}|)+\sigma(|Y_{f}|<|Y_{{\bar
f}}|)}\right|_{|p_{z,{f\bar f}}|>p^{\rm{cut}}_{Z,{f\bar f}}},
 \end{equation}
\begin{equation}
A_{\rm C}(Y_{\rm C}) = \frac{\sigma_{ f}(|Y_{f}|<Y_{\rm
C})-\sigma_{{\bar f}}(|Y_{{\bar f}}|<Y_{\rm C})} {\sigma_{
f}(|Y_{f}|<Y_{\rm C})+\sigma_{{\bar f}}(|Y_{{\bar f}}|<Y_{\rm C})},
 \end{equation}
\begin{equation}
A_{\rm E}(Y_{\rm C}) = \frac{\sigma_{f}(Y_{\rm C}<|Y_{
f}|)-\sigma_{{\bar f}}(Y_{\rm C}<|Y_{{\bar f}}|)} {\sigma_{
f}(Y_{\rm C}<|Y_{f}|)+\sigma_{{\bar f}}(Y_{\rm C}<|Y_{{\bar f}}|)}
 \end{equation}
in which $Y$ is the rapidity of ${f}/{\bar f}$ or fermion pair
accordingly. The $p_{z,{f\bar f}}$ is the z-direction momentum of
the fermion pair.

In order to keep the partonic asymmetry even at the hadronic level,
no matter how different these asymmetries look like,
each of them has used the fact that
the energy fraction of the valence quark is usually larger than that
of the sea quark in the proton. These asymmetries can be
classified into two categories according to their similarities.
$\rm{A_C}$ and $\rm{A_E}$ belong to one category and the remaining
two belong to the other. For $\rm{A_C}$ and $\rm{A_E}$, they account
two complementary kinematical regions, namely central region $|Y_{{f, \bar
f}}|<Y_{\rm C}$ and the edge region $|Y_{{f, \bar f}}|>Y_{\rm C}$ in
the laboratory frame respectively. $\rm{A_E}$ can usually suppress more
efficiently the symmetric $gg\to q\bar{q}$ background events which
mostly distributes in the central region than that of the $\rm{A_C}$
\cite{arXiv.1101.2507}. Therefore $\rm{A_E}$ is usually more significant than the
$\rm{A_C}$. The difference between $\rm{A_{RFB}}$ and $\rm{A_{OFB}}$
is the cuts in order to keep the partonic asymmetry.
$Y_{{f\bar f}}$ and $P_{{f\bar f}}^z$ are proportional to $(x_1-x_2)/(x_1+x_2)$ and
$(x_1-x_2)$ respectively,
where $x_1$ and $x_2$ are the momentum fraction of the two colliding
partons. The most important difference between the two categories
is that the asymmetry utilizes the different kinematical information.
The asymmetries in the first category utilize either ${f}$ or ${\bar{f}}$ momentum information.
While the asymmetries of the second category require to
measure the kinematical information of $f$ and $\bar f$ simultaneously.

Due to their different characteristics, the four types of
asymmetries will be used in different cases. Each of them have their most suitable places. In the
followings, we will investigate how to use four asymmetries
to discriminate different $Z^\prime$ models, namely which one is the most suitable type of
asymmetry and the corresponding optimal cuts.

\section{Discriminating different $Z^\prime$s via asymmetries}\label{OPTIMIZATION}

\subsection{$Z^\prime$ in the left-right model and sequential standard model}

In this paper in order to illustrate how to utilize asymmetries to discriminate
different $Z^\prime$s, we choose two test models as the examples: left-right model and
sequential standard model.

The left-right model is based on the symmetry group $SU(2)_R \times
SU(2)_L\times U(1)_{B-L}$, where $B-L$ is the difference between
baryon and lepton numbers. The couplings between the $Z^\prime$ and
fermions are \cite{PhysLettB.583.111, PhysRevD.77.115004}
\begin{equation}
g_{Z^\prime}J^{\mu}_{Z^\prime}Z^\prime_{\mu}=\frac{e}{c_W}\sum\limits_f\bar\psi_f\gamma^\mu
\left[\frac{1-\gamma^5}{2}g_L^{fZ^\prime}+\frac{1+\gamma^5}{2}g_R^{fZ^\prime}\right]\psi_fZ^\prime_\mu.
\end{equation}

As the test model, the sequential standard model $Z^\prime$ has the same fermion couplings as
the SM $Z$ boson and which can be written as,
\begin{equation}
g_{Z^\prime}J^{\mu}_{Z^\prime}Z^\prime_{\mu}=-\frac{g}{2c_W}
\sum\limits_f\bar\psi_f\gamma_\mu\left(g_V^f-g_A^f\gamma_5\right)\psi_fZ^\prime_\mu.
\end{equation}

The parameters in these two models can be summarized in Tab. \ref{coupling}.
\begin{table}[htb]
\caption{Couplings of the ${Z^\prime}$ boson to the SM fermions in the
left-right model and the sequential SM. $\alpha_{LR}=\sqrt{(c_W^2
g_R^2/s_W^2 g_L^2)-1}$, where $g_L=e/s_W$ and $g_R$ are the
$SU(2)_L$ and $SU(2)_R$ coupling constants with $s_W^2=1-c_W^2\equiv
\sin^2\theta_W$\cite{PhysLettB.583.111}.}
\begin{tabular}{|c|c|c|c|c|}
\hline\hline $f$ & $g_L^{f{Z^\prime}}$ & $g_R^{f{Z^\prime}}$& $g_V^f$ & $g_A^f$\\
\hline $e$ & $\frac{1}{2\alpha_{LR}}$
&$\frac{1}{2\alpha_{LR}}-\frac{\alpha_{LR}}{2}$ & $ -\frac{1}{2}+2\sin^2\theta_W$ & $-\frac{1}{2}$\\
$u$ & $-\frac{1}{6\alpha_{LR}}$ &
$-\frac{1}{6\alpha_{LR}}+\frac{\alpha_{LR}}{2}$& $\frac{1}{2}-\frac{4}{3}\sin^2\theta_W$ & $\frac{1}{2}$ \\
$d$  & $-\frac{1}{6\alpha_{LR}}$ &
$-\frac{1}{6\alpha_{LR}}-\frac{\alpha_{LR}}{2}$&
$-\frac{1}{2}+\frac{2}{3}\sin^2\theta_W$ &
$-\frac{1}{2}$\\\hline\hline
\end{tabular}
\label{coupling}
\end{table}
Throughout the paper the mass
of ${Z^\prime}$ is set to be 1.5 TeV for both models. The
$\alpha_{\rm{LR}}$ in LR model is set to be 1.88 as the benchmark point, in order
to make the width of ${Z^\prime}$ in the LR model the same as that in the SSM.
SM parameters are chosen as
$\alpha=1/127.9$, $\sin^2\theta_W=0.231$, $M_{\rm Z}=91.133\rm{GeV}$,
$\Gamma_Z=2.495 \mbox{GeV}$ and  $m_t=171.2\mbox{GeV}$.
In calculating the width of the ${Z^\prime}$, only its decays to
the SM fermions are included.

\subsection{Optimizing asymmetry}

In our analysis, the basic kinematical cuts are taken as  $p_T>20\rm{GeV}$ and $Y<2.5$
for the leptons, $p_T>0.3M_{\rm {Z^\prime}}$ and $Y<2.5$ for the bottom
and top quarks. Here the $p_T$ cut for quarks can suppress the QCD
backgrounds.  The LHC energy $\sqrt{s}$ is set to be 14 TeV.

\begin{figure}[htbp]

\begin{center}
\includegraphics[width=0.40\textwidth]
{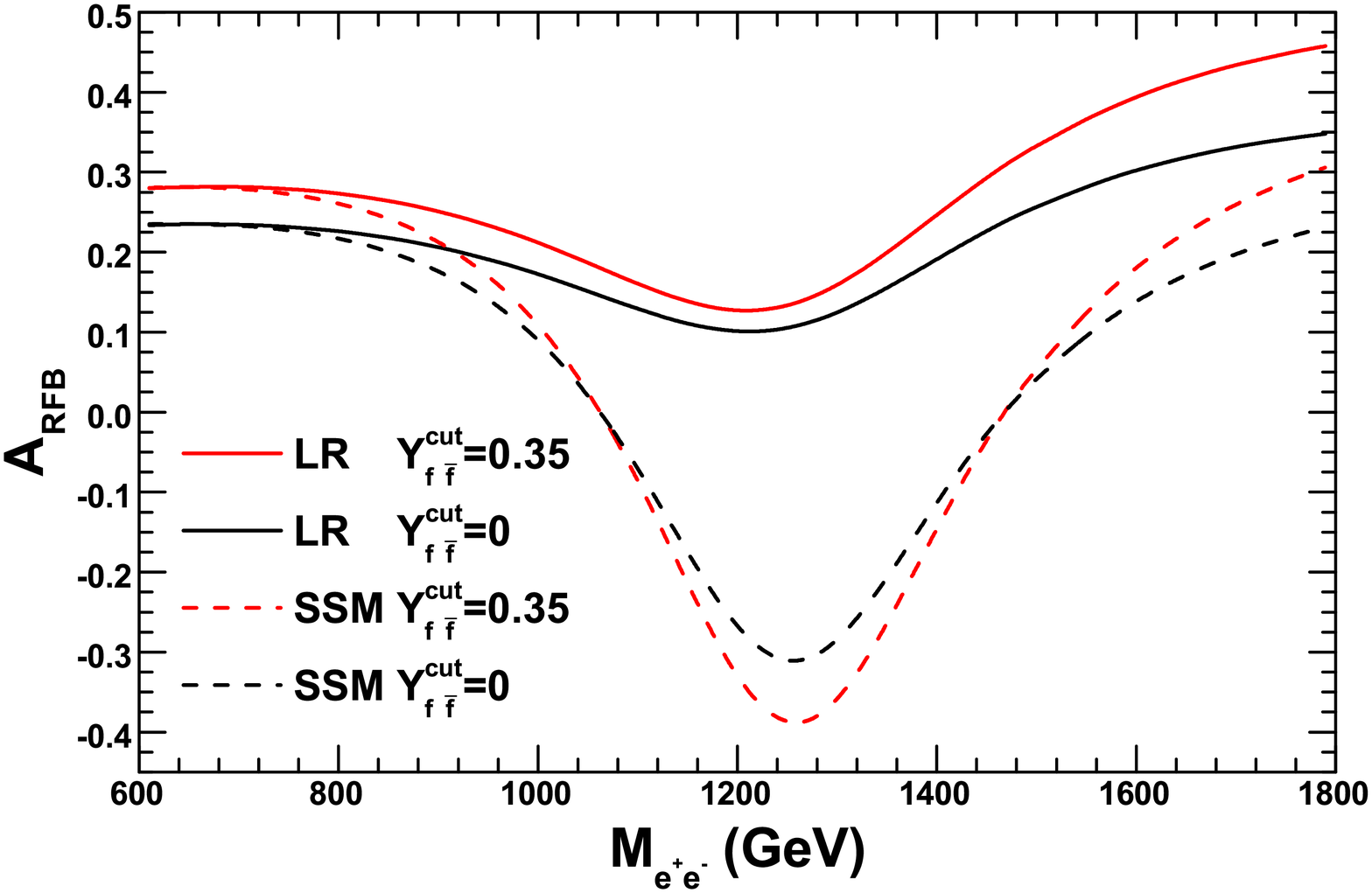}
\end{center}

\caption{\label{eedAdM}  $A_{\rm{RFB}}$ for the process $pp\rightarrow e^+e^- X$
as a function of $M_{e^+e^-}$ in LR mode and SSM with cuts $Y_{{f\bar f}}^{\rm{cut}}=0$ and
$Y_{{f\bar f}}^{\rm{cut}}=0.35$ respectively.}
\end{figure}

Fig. \ref{eedAdM} shows $A_{\rm{RFB}}$ for the process $pp\rightarrow e^+e^- X$
as a function of $M_{e^+e^-}$. From the figure it is clear that the asymmetry depends on $M_{e^+e^-}$. As depicted above, the asymmetry depends on the chiral properties of $Z^\prime$ and SM fermions, as
well as the selected data sample. In order to keep the asymmetry information as much as possible,
in our analysis the four data samples are chosen, same with those
in Ref. \cite{arXiv.0910.1334}. They are the on-peak events with
$M_{\rm{{Z^\prime}}}-3\Gamma < M_{\mu^+\mu^-(e^+e^-)} < M_{\rm{{Z^\prime}}}+3\Gamma$,
off-peak events with $2/3M_{\rm{{Z^\prime}}}< M_{\mu^+\mu^-(e^+e^-)} < M_{\rm{{Z^\prime}}}-3\Gamma$,
on-peak events
with $M_{\rm{{Z^\prime}}}-2.5\Gamma < M_{b\bar b} <M_{\rm{{Z^\prime}}}+2.5\Gamma$ and
on-peak events with $M_{\rm{{Z^\prime}}}-2.5\Gamma< M_{t\bar t}< M_{\rm{{Z^\prime}}}+2.5\Gamma$.

From Fig. \ref{eedAdM} $A_{\rm{RFB}}$ with $Y_{{f\bar f}}^{\rm{cut}}=0.35$ is larger than that
with $Y_{{f\bar f}}^{\rm{cut}}=0$. However the magnitude of $A_{\rm{RFB}}$ is not a good measure
to optimize the observable. As usual we utilize
the significance of the
asymmetry as a measure to select optimal cuts. The significance is defined as
\begin{eqnarray}
S_{A}\equiv \frac{\sigma^A \mathcal{L}}{\sqrt{\sigma
\mathcal{L}}}=A_{\rm{FB}}\sqrt{\mathcal{L}\ \sigma},
\end{eqnarray}
where $A_{\rm{FB}}$ can be $A_{\rm{RFB}}$, $A_{\rm{OFB}}$, $A_{\rm
C}$ or $A_{\rm E}$, $\mathcal{L}$ is the LHC integrated luminosity
which is taken as $100\rm{fb}^{-1}$ throughout our analysis, and
$\sigma$ is the cross section. In this part, the detection efficiency is set to be 1.

\begin{figure}[htbp]

\begin{center}
\includegraphics[width=0.40\textwidth]
{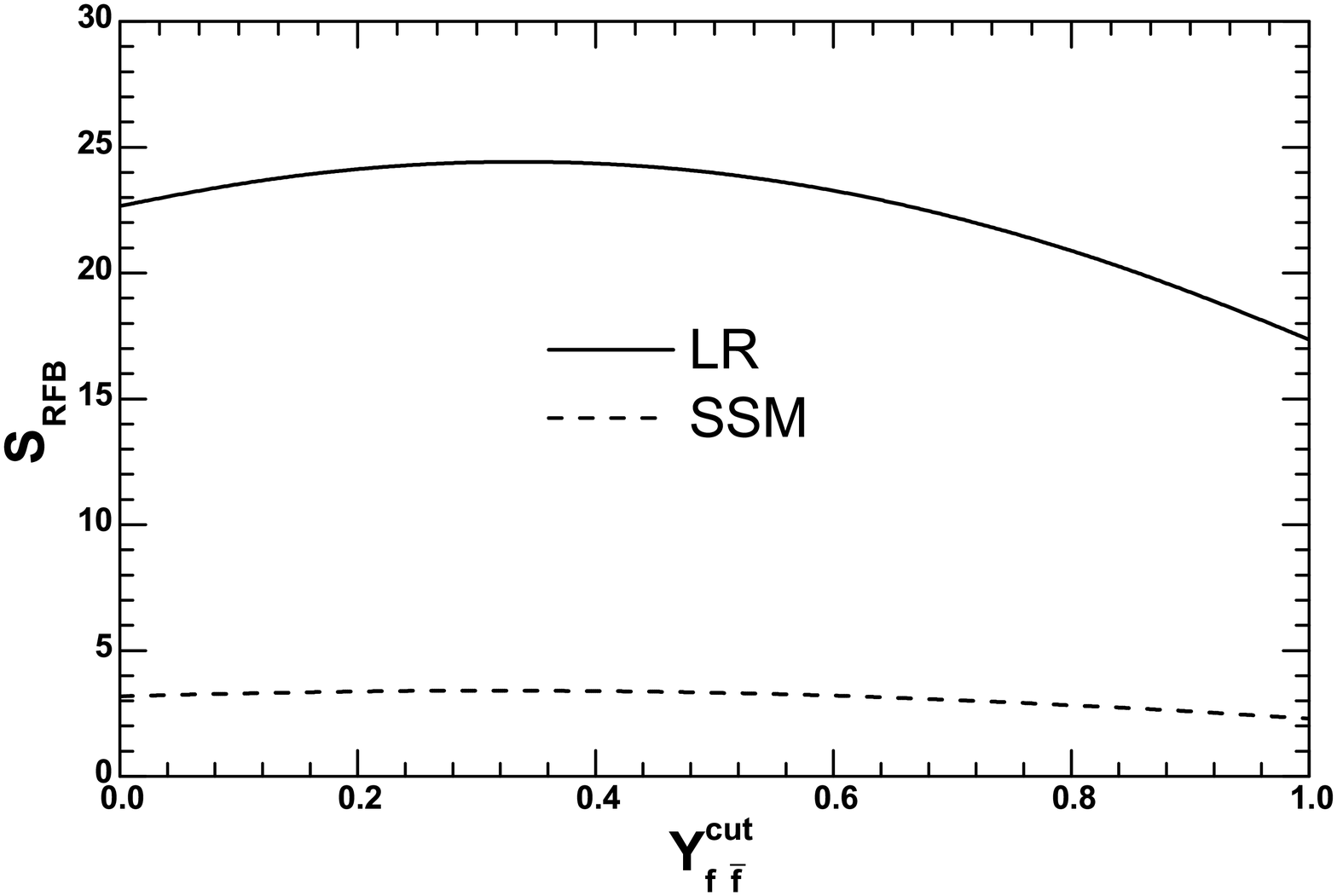}
\includegraphics[width=0.40\textwidth]
{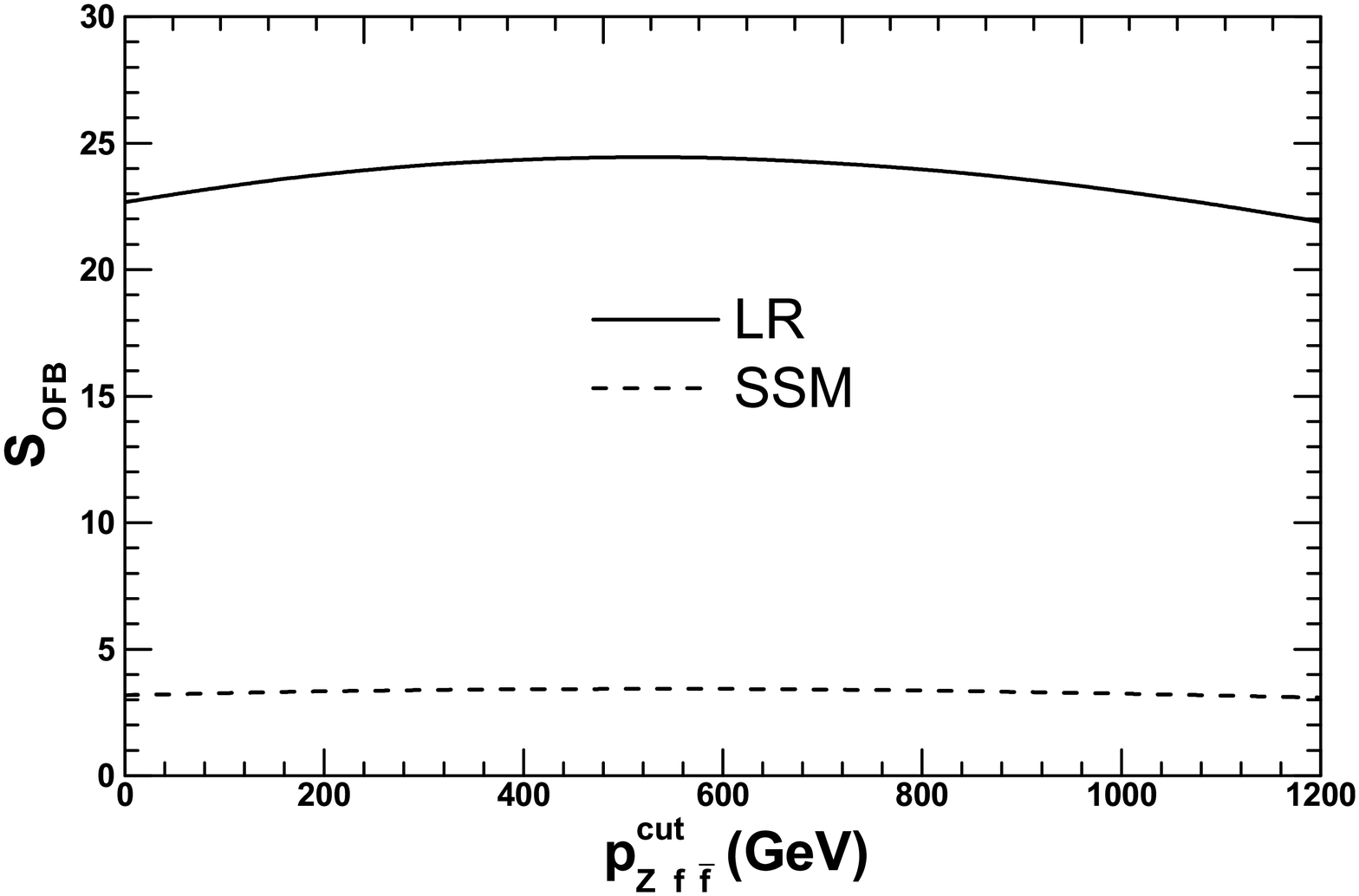}
\end{center}
\begin{center}
\includegraphics[width=0.40\textwidth]
{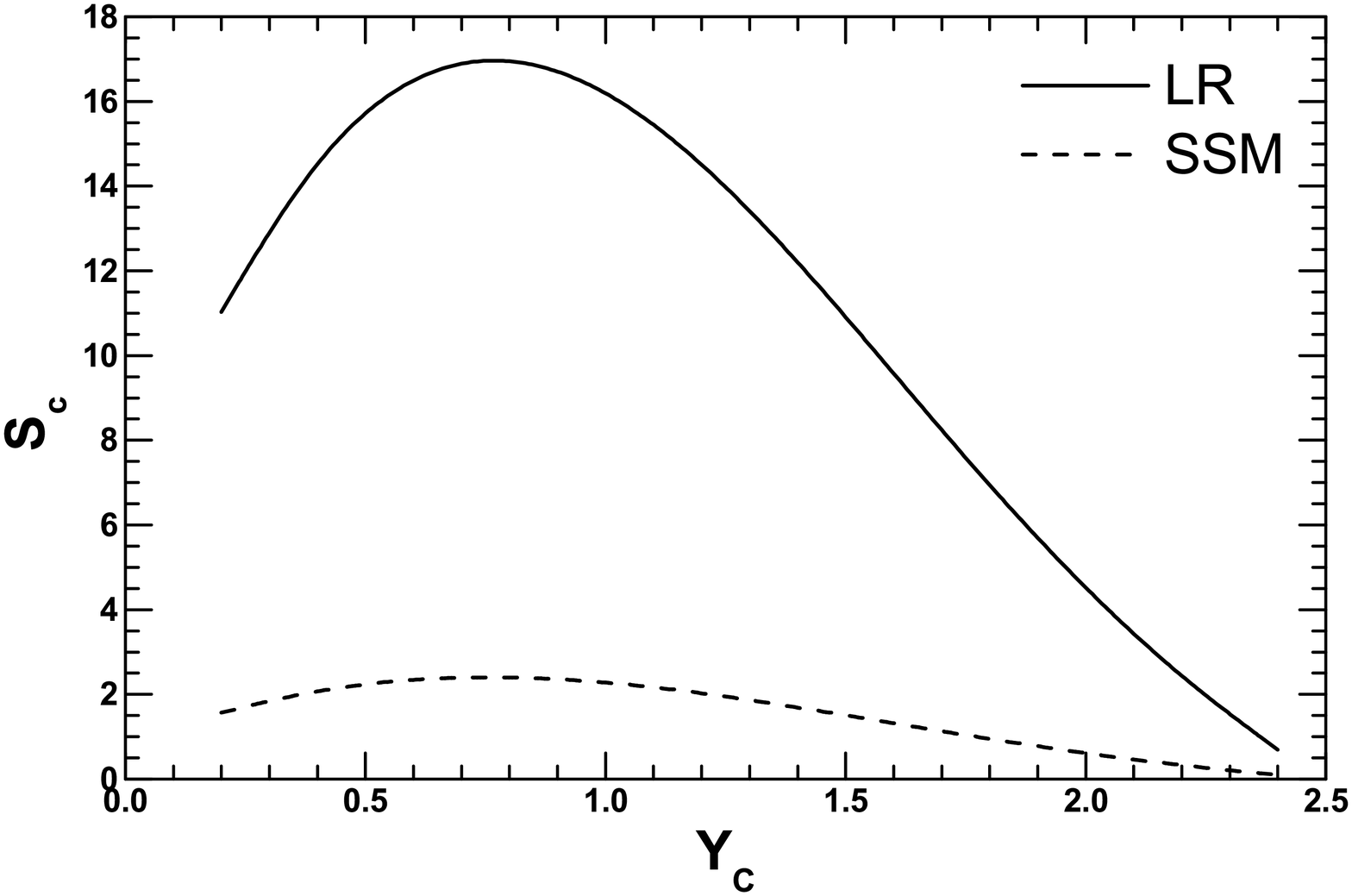}
\includegraphics[width=0.40\textwidth]
{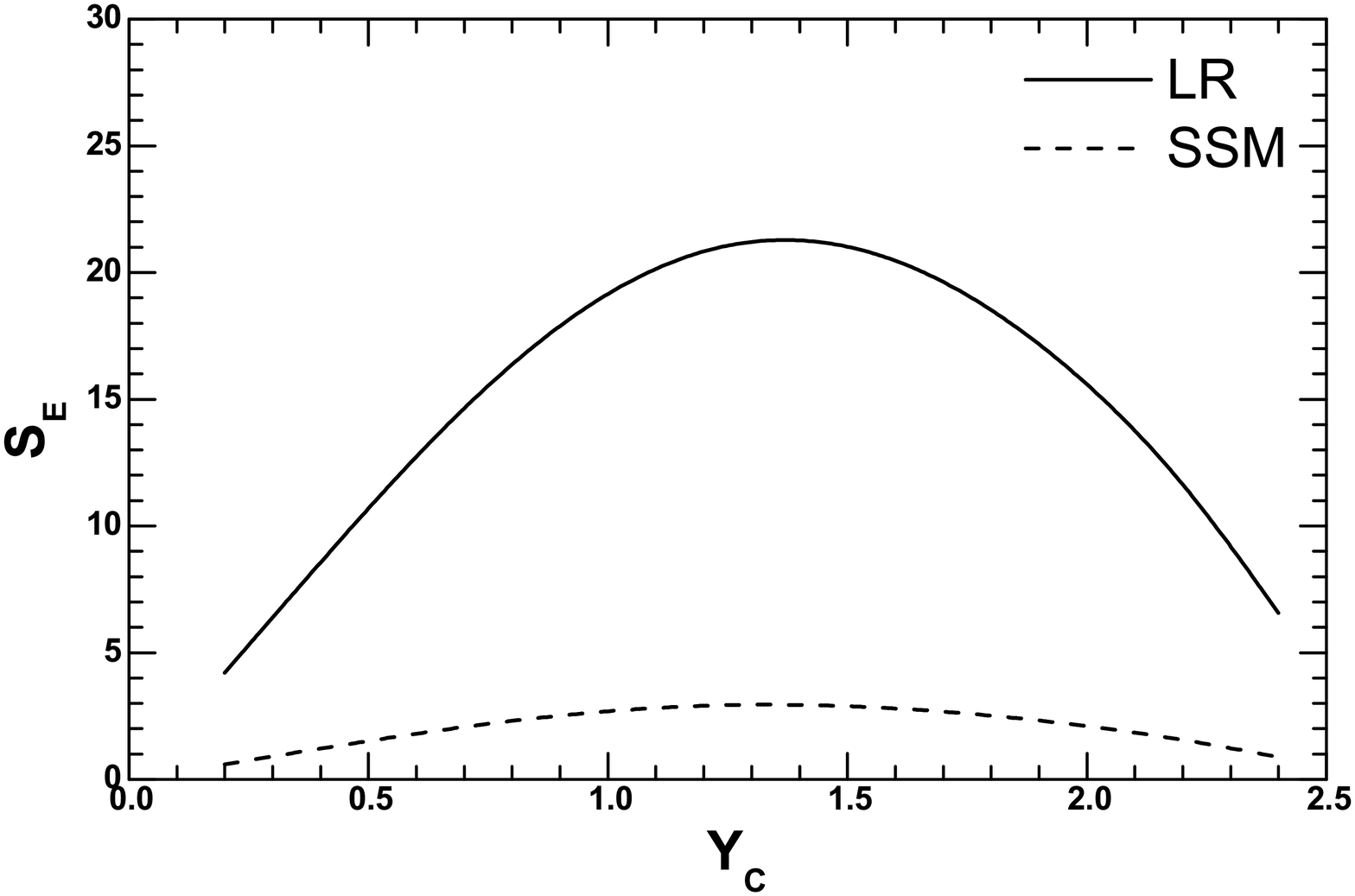}
\end{center}

\caption{\label{eeAsymmetrycut} Significance as a function of corresponding cut for on-peak $e^+e^-$ events
 in LR model and SSM.}

\end{figure}

\begin{figure}[htbp]

\begin{center}
\includegraphics[width=0.40\textwidth]
{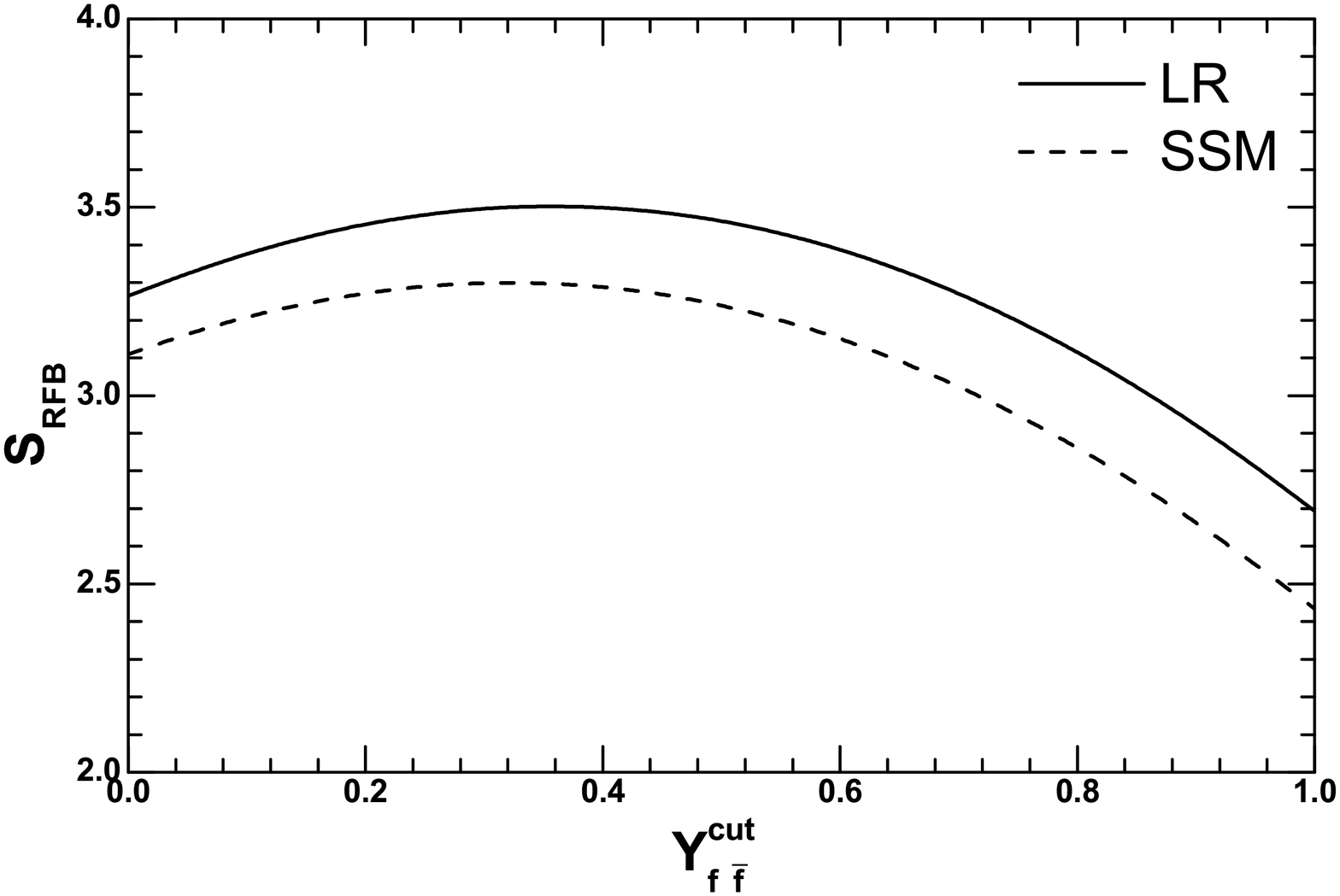}
\includegraphics[width=0.40\textwidth]
{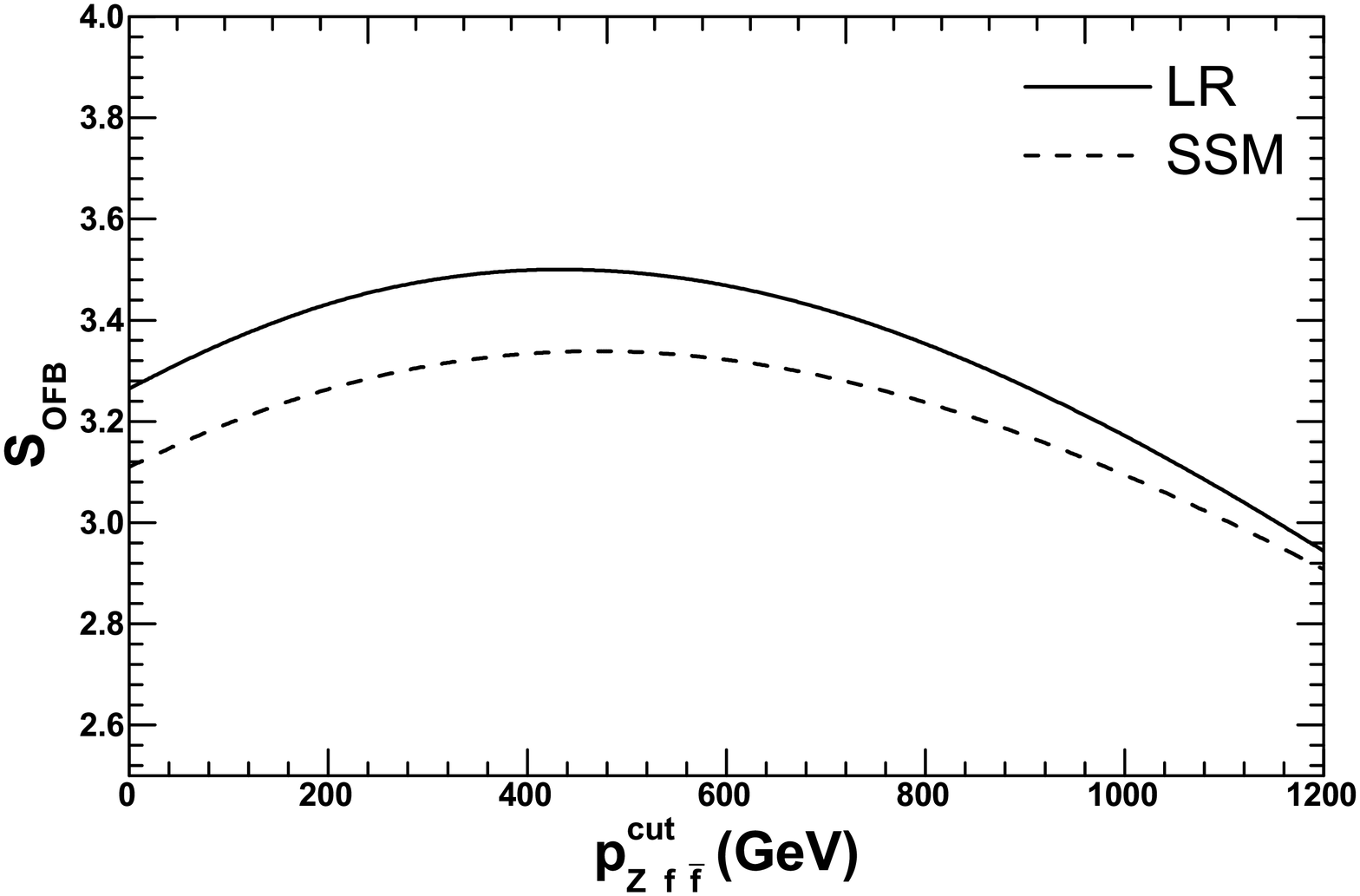}
\end{center}
\begin{center}
\includegraphics[width=0.40\textwidth]
{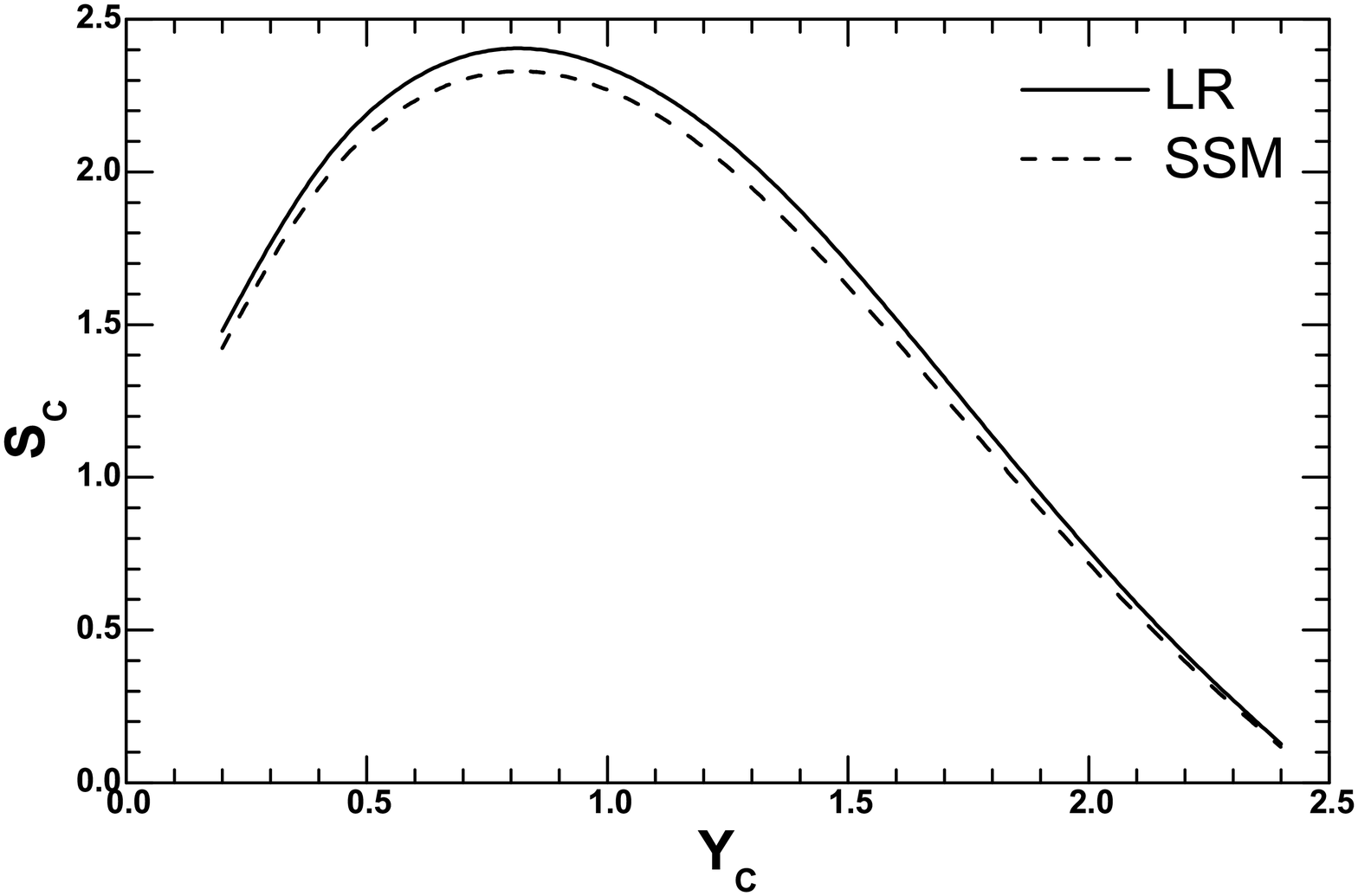}
\includegraphics[width=0.40\textwidth]
{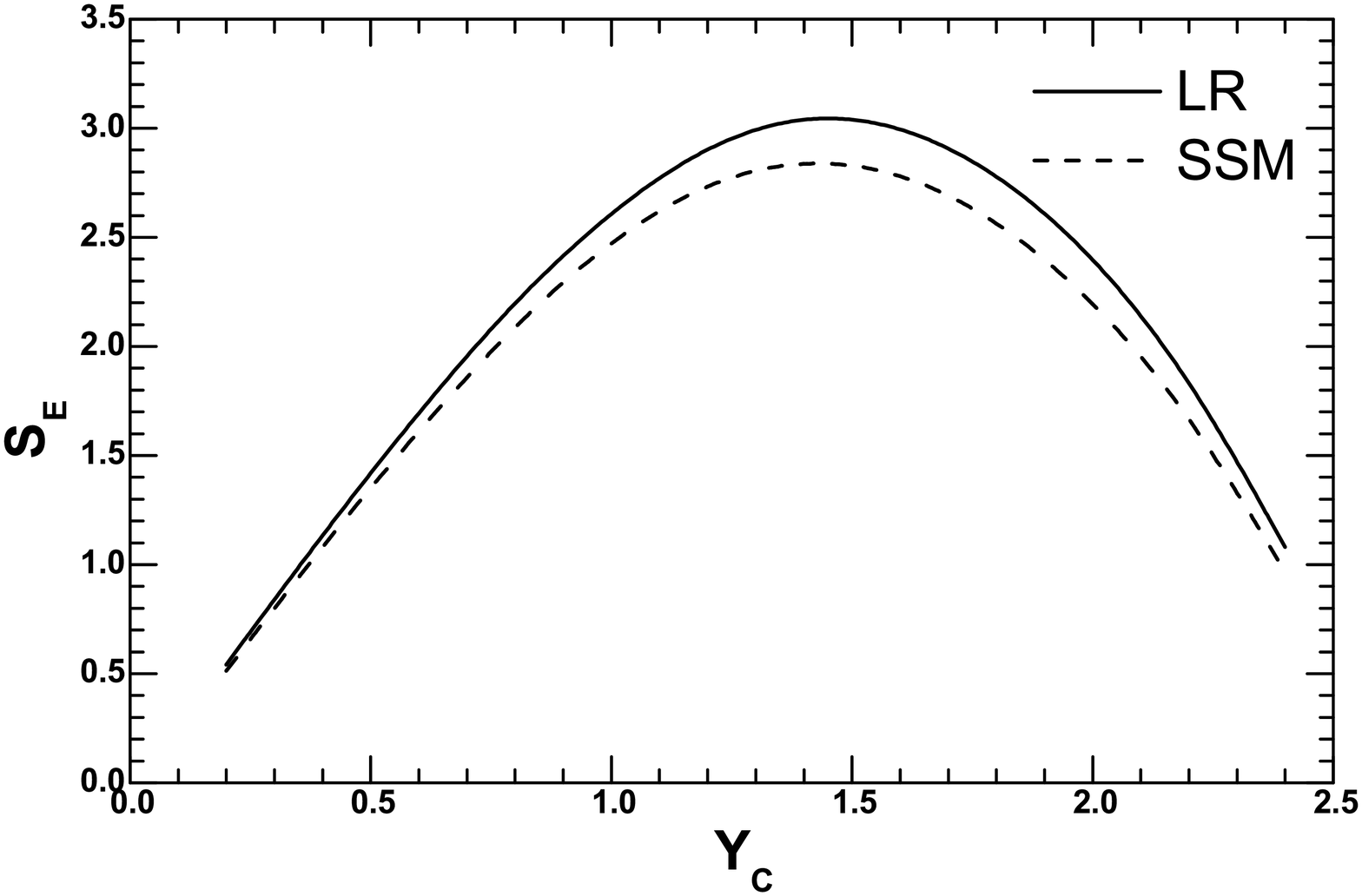}
\end{center}

\caption{\label{eeAsymmetrycutOffshell} Same with Fig. \ref{eeAsymmetrycut} except for off-peak $e^+e^-$ events.}

\end{figure}

\begin{figure}[htbp]

\begin{center}
\includegraphics[width=0.40\textwidth]
{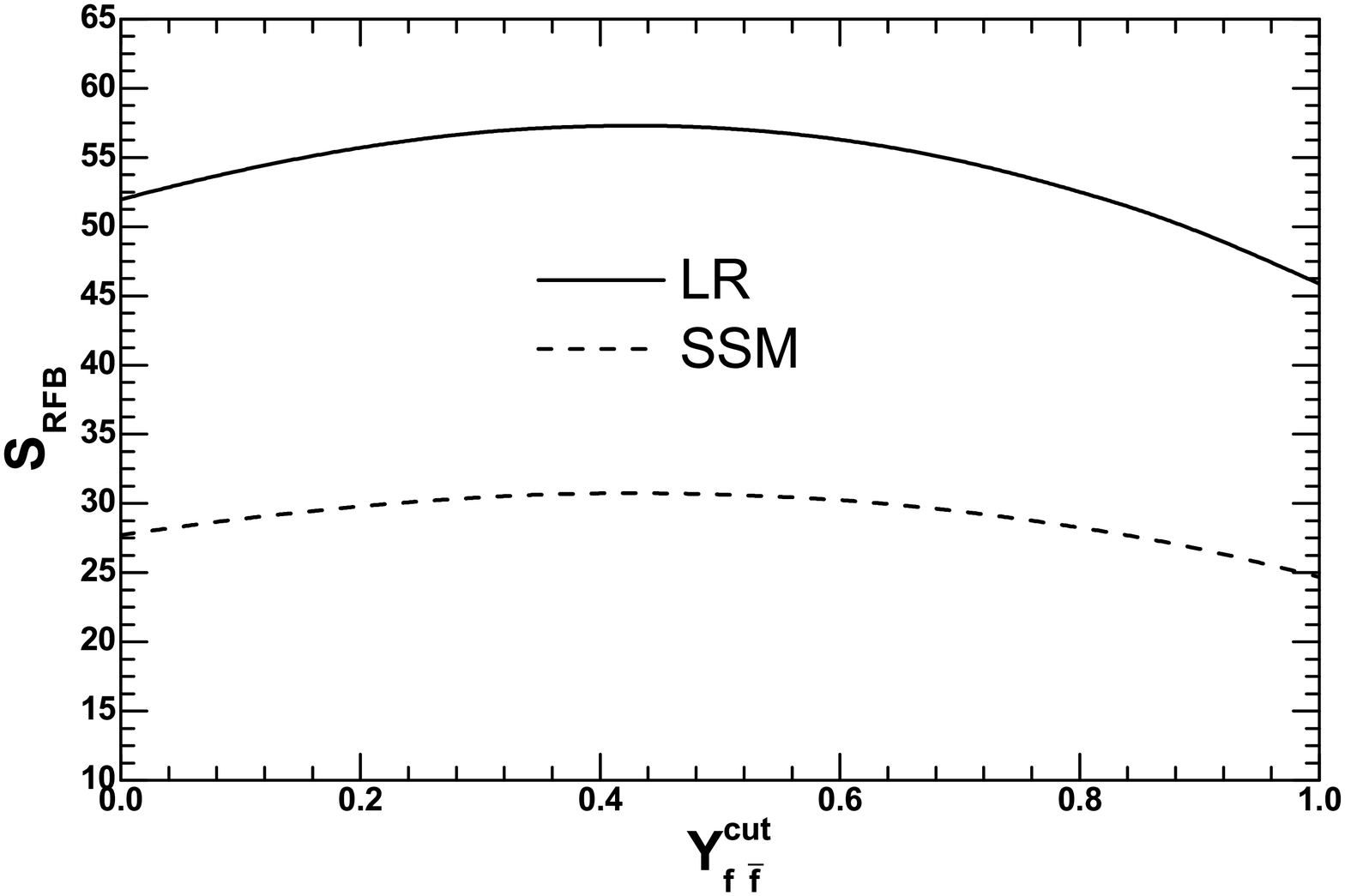}
\includegraphics[width=0.40\textwidth]
{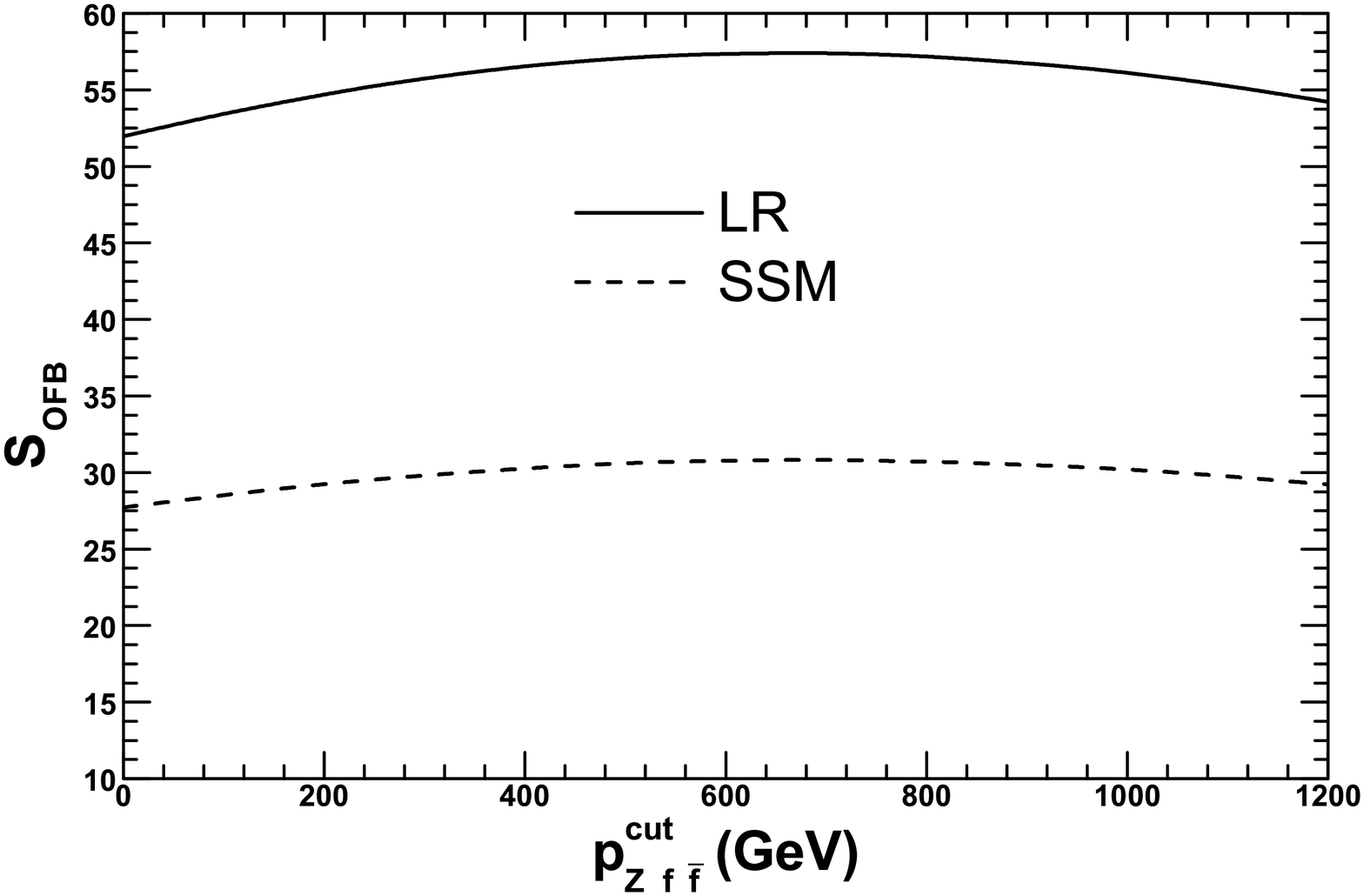}
\end{center}
\begin{center}
\includegraphics[width=0.40\textwidth]
{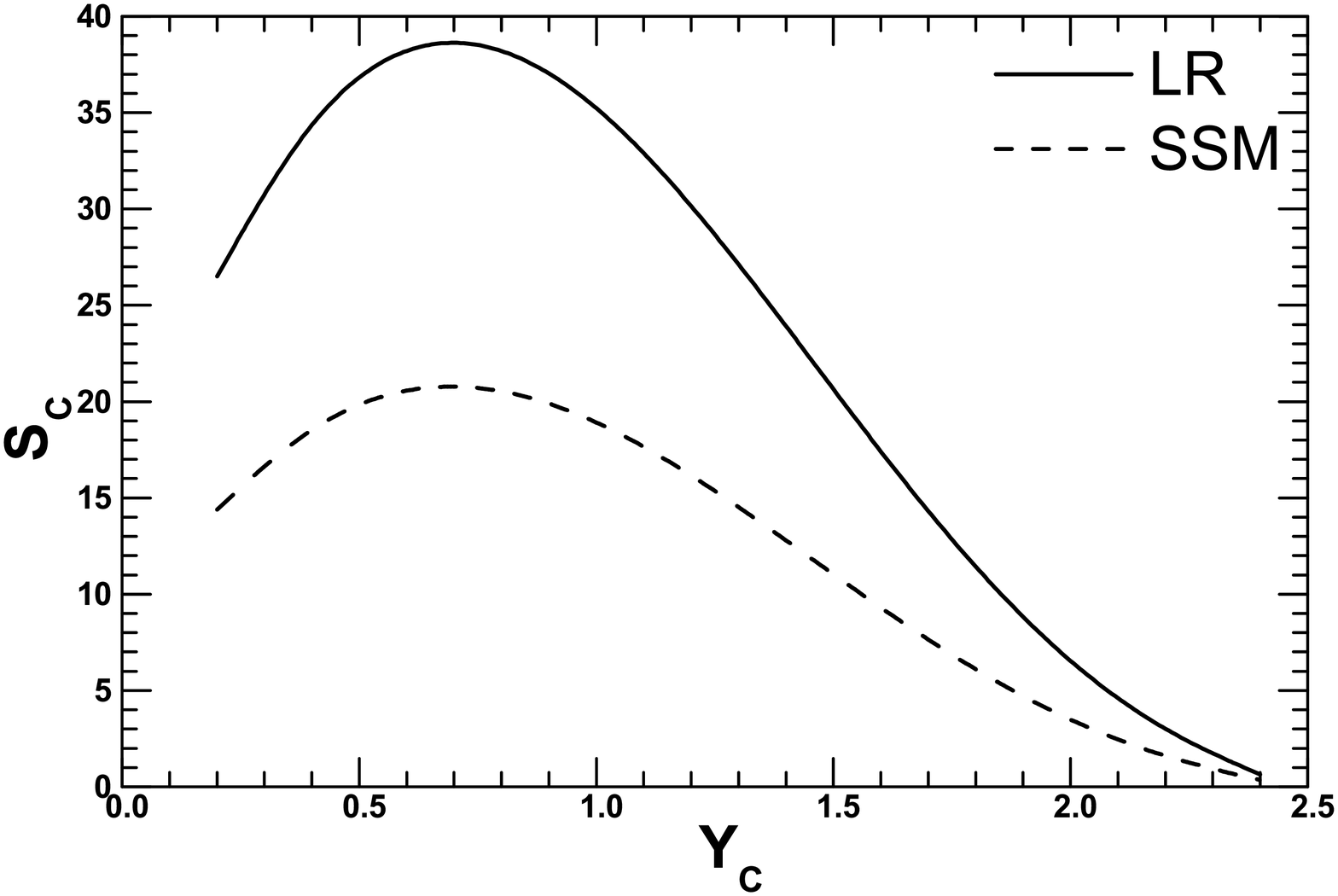}
\includegraphics[width=0.40\textwidth]
{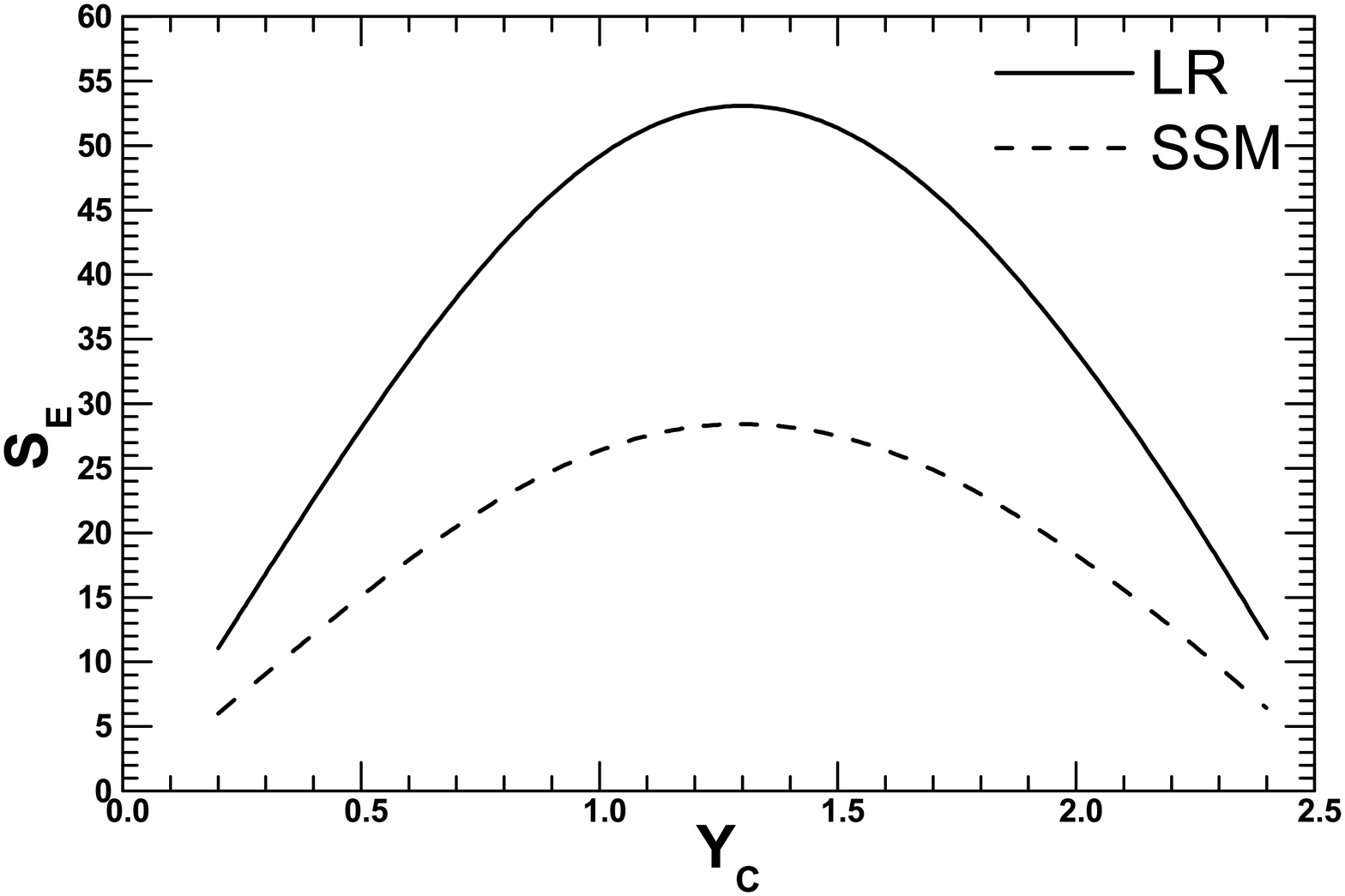}
\end{center}

\caption{\label{bbAsymmetrycut}Same with Fig. \ref{eeAsymmetrycut} except for
on-peak $b\bar b$ events. }

\end{figure}

\begin{figure}[htbp]
\begin{center}
\includegraphics[width=0.40\textwidth]
{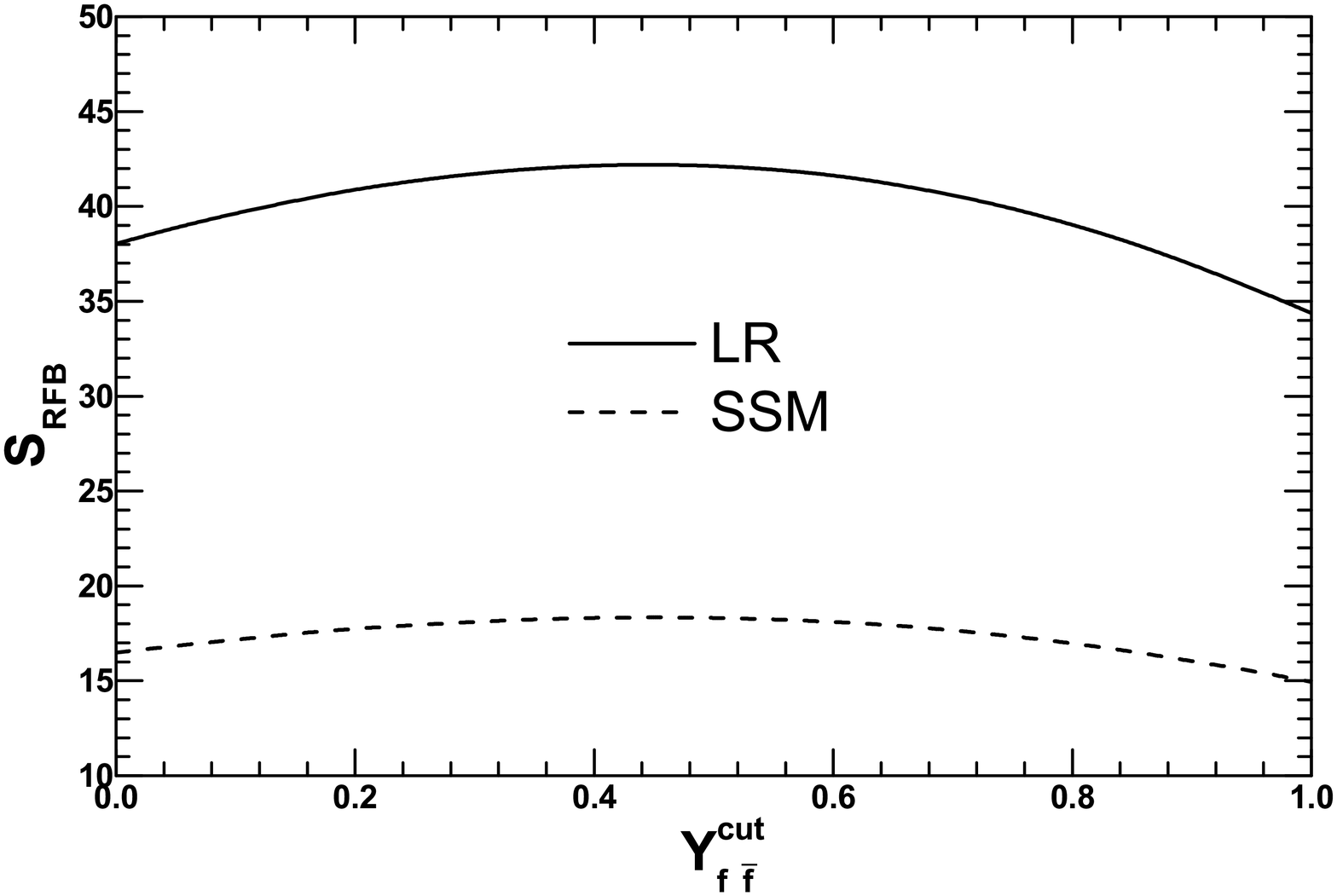}
\includegraphics[width=0.40\textwidth]
{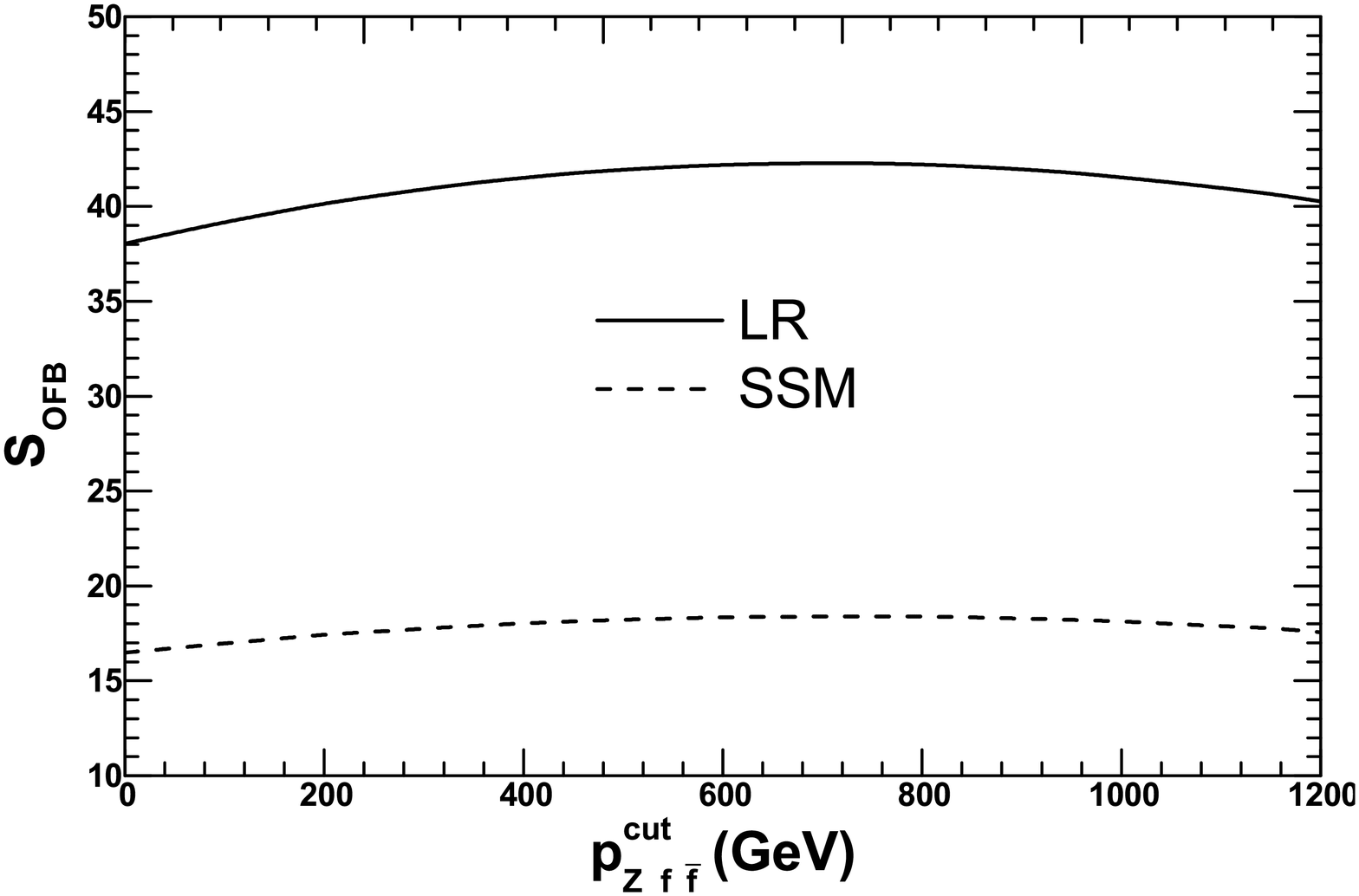}
\end{center}
\begin{center}
\includegraphics[width=0.40\textwidth]
{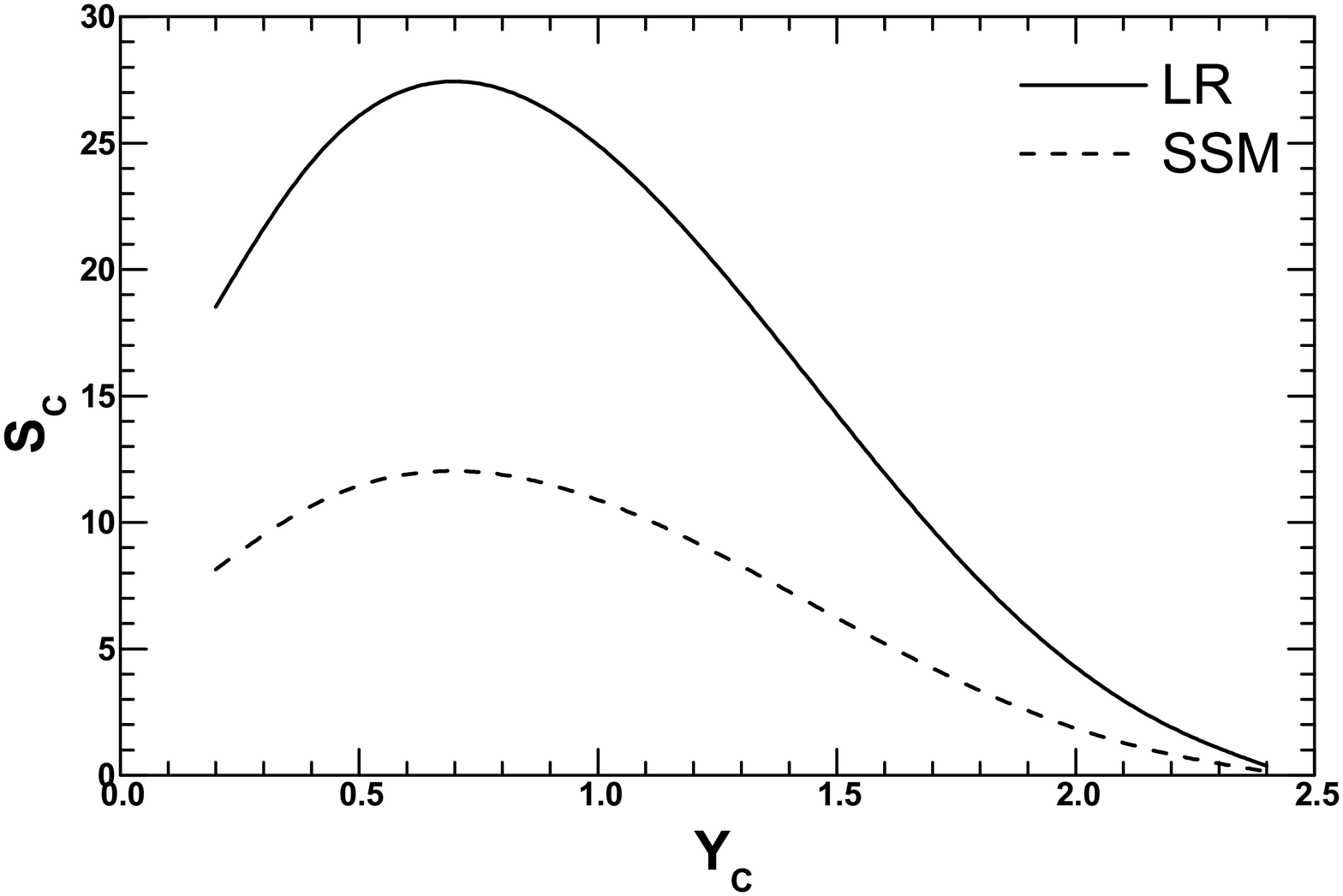}
\includegraphics[width=0.40\textwidth]
{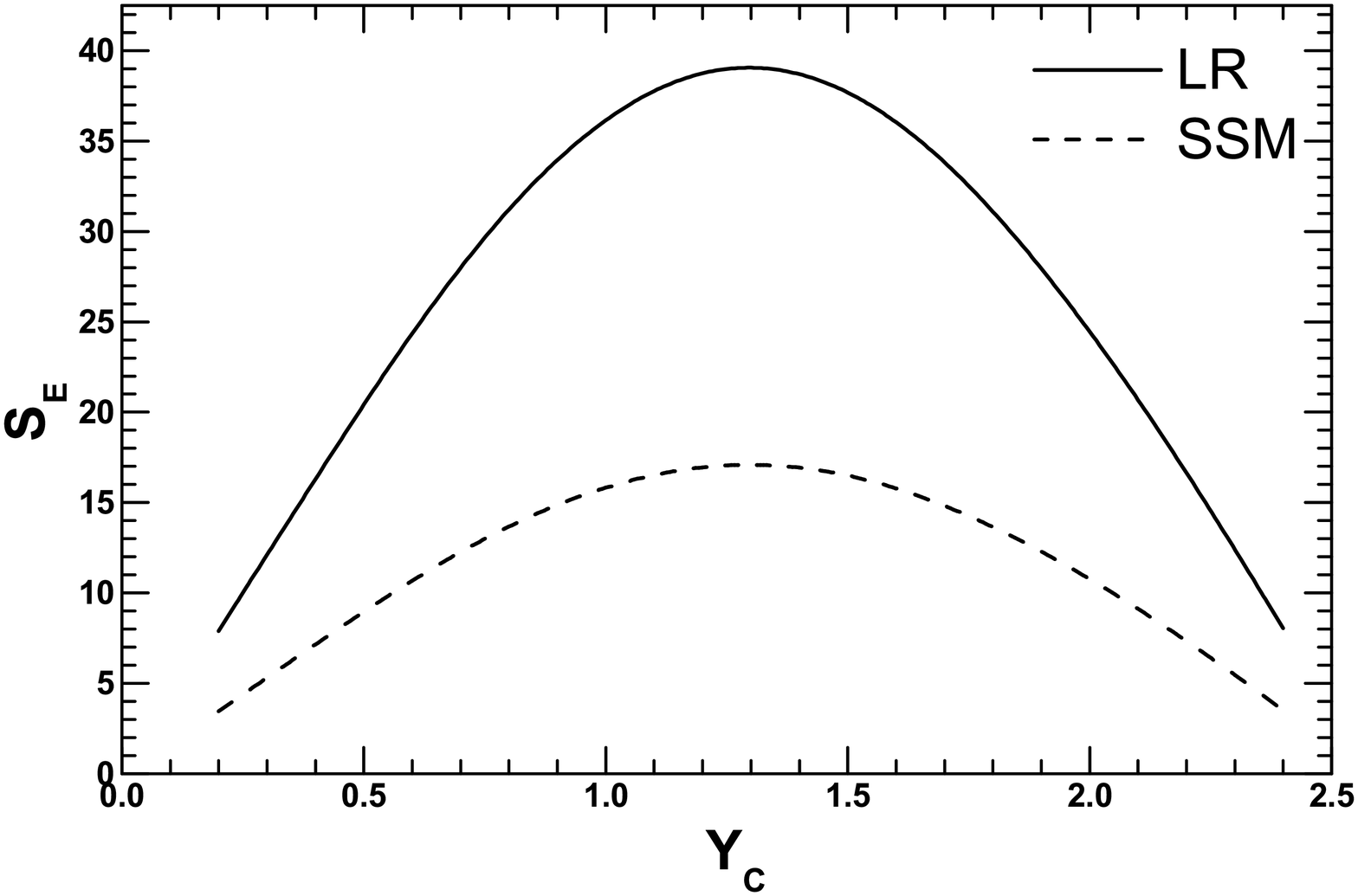}
\end{center}
\caption{\label{ttAsymmetrycut}Same with Fig. \ref{eeAsymmetrycut} except for
on-peak $t\bar t$ events.}
\end{figure}

Figs. \ref{eeAsymmetrycut}, \ref{eeAsymmetrycutOffshell},
\ref{bbAsymmetrycut} and \ref{ttAsymmetrycut} show the
significance as a function of the corresponding cut for four asymmetries and four
data samples respectively. In order to achieve the maximum significance the corresponding best cuts are depicted in Tabs. \ref{ee-optimized},
\ref{eeOffshell-optimized}, \ref{bb-optimized} and \ref{tt-optimized} respectively.

\begin{table}[htb]
\caption{\label{ee-optimized}Optimized cut and the corresponding maximum
value of significance for on-peak $e^+e^-$ events for the four kinds
of asymmetries in both LR model and SSM.}

\begin{tabular}{|c|}
\hline LR
\\
\hline
\begin{tabular}{c|c|c|c|c}
 & $\rm{A_{RFB}}$ & $\rm{A_{OFB}}$ & $\rm{A_{C}}$ & $\rm{A_{E}}$
\\
\hline
 Best cut & $Y^{{f\bar f}}_{\rm{cut}}=0.35$ & $P^{{f\bar
 f}}_{\rm{z,cut}}=550\rm{GeV}$ & $Y_{\rm C}=0.8$ & $Y_{\rm C}=1.4$
\\
\hline
 Significance(with $100\rm{fb}^{-1}$) & 24.4 & 24.4 & 17.1 & 21.5
\end{tabular}
\\
\hline  SSM
\\
\hline
\begin{tabular}{c|c|c|c|c}
 & $\rm{A_{RFB}}$ & $\rm{A_{OFB}}$ & $\rm{A_{C}}$ & $\rm{A_{E}}$
\\
\hline
 Best cut & $Y^{{f\bar f}}_{\rm{cut}}=0.35$ &  $P^{{f\bar
 f}}_{\rm{z,cut}}=550\rm{GeV}$ & $Y_{\rm C}=0.8$ & $Y_{\rm C}=1.4$
\\
\hline
 Significance(with $100\rm{fb}^{-1}$) &
 3.40 & 3.43 & 2.43 & 2.97  \\ \hline
\end{tabular}
\end{tabular}

\end{table}

\begin{table}[htb]
\caption{\label{eeOffshell-optimized} Same with Tab. \ref{ee-optimized} except for off-peak $e^+e^-$ events.}
\begin{tabular}{|c|}
\hline LR \\ \hline
\begin{tabular}{c|c|c|c|c}
 &$\rm{A_{RFB}}$ & $\rm{A_{OFB}}$ & $\rm{A_{C}}$ & $\rm{A_{E}}$  \\
\hline
 Best cut & $Y^{{f\bar f}}_{\rm{cut}}=0.35$ & $P^{\rm{f\bar
 f}}_{\rm{z,cut}}=450\rm{GeV}$ & $Y_{\rm C}=0.8$ & $Y_{\rm C}=1.4$
\\
\hline
 Significance(with $100\rm{fb}^{-1}$) & 3.51 & 3.51 & 2.43 & 3.07
\end{tabular}
\end{tabular}
\begin{tabular}{|c|}
\hline SSM
\\ \hline
\begin{tabular}{c|c|c|c|c}
 &$\rm{A_{RFB}}$ & $\rm{A_{OFB}}$ & $\rm{A_{C}}$ & $\rm{A_{E}}$
\\
\hline
 Best cut & $Y^{{f\bar f}}_{\rm{cut}}=0.35$ & $P^{\rm{f\bar
 f}}_{\rm{z,cut}}=450\rm{GeV}$ & $Y_{\rm C}=0.8$ & $Y_{\rm C}=1.4$
\\
\hline
 Significance(with $100\rm{fb}^{-1}$) & 3.30 & 3.34 & 2.36 & 2.87  \\
 \hline
\end{tabular}
\end{tabular}
\end{table}

\begin{table}[htb]
\caption{\label{bb-optimized}Same with Tab. \ref{ee-optimized} except for on-peak $b\bar b$ events.}

\begin{tabular}{|c|}
\hline LR
\\
\hline
\begin{tabular}{c|c|c|c|c}
 &$\rm{A_{RFB}}$& $\rm{A_{OFB}}$ & $\rm{A_{C}}$ & $\rm{A_{E}}$
\\
\hline
 Best cut & $Y^{{f\bar f}}_{\rm{cut}}=0.45$ & $P^{\rm{f\bar
 f}}_{\rm{z,cut}}=700\rm{GeV}$ & $Y_{\rm C}=0.6$ & $Y_{\rm C}=1.2$
\\
\hline
 Significance(with $100\rm{fb}^{-1}$) & 57.2 & 57.4 & 38.8 & 53.1
\end{tabular}
\\
\hline SSM
\\
\hline
\begin{tabular}{c|c|c|c|c}
 &$\rm{A_{RFB}}$ & $\rm{A_{OFB}}$ & $\rm{A_{C}}$ & $\rm{A_{E}}$
\\
\hline
 Best cut & $Y^{{f\bar f}}_{\rm{cut}}=0.45$& $P^{\rm{f\bar
 f}}_{\rm{z,cut}}=700\rm{GeV}$ & $Y_{\rm C}=0.6$ & $Y_{\rm C}=1.2$
\\
\hline
 Significance(with $100\rm{fb}^{-1}$) & 30.7 & 30.8 & 20.9 & 28.5
\end{tabular}
\\
\hline
\end{tabular}

\end{table}

\begin{table}[htb]
\caption{\label{tt-optimized}Same with Tab. \ref{ee-optimized} except for on-peak $t\bar t$ events.}

\begin{tabular}{|c|}
\hline  LR
\\
\hline
\begin{tabular}{c|c|c|c|c}
 &$\rm{A_{RFB}}$ & $\rm{A_{OFB}}$& $\rm{A_{C}}$ & $\rm{A_{E}}$
\\
\hline
 Best cut & $Y^{{f\bar f}}_{\rm{cut}}=0.45$ & $P^{\rm{f\bar
 f}}_{\rm{z,cut}}=700\rm{GeV}$ & $Y_{\rm C}=0.6$ & $Y_{\rm C}=1.2$
\\
\hline
 Significance(with $100\rm{fb}^{-1}$) & 42.2 & 42.3 & 27.6 & 39.2
\end{tabular}
\\
\hline SSM
\\
\hline
\begin{tabular}{c|c|c|c|c}
 &$\rm{A_{RFB}}$& $\rm{A_{OFB}}$ & $\rm{A_{C}}$ & $\rm{A_{E}}$
\\
\hline
 Best cut & $Y^{{f\bar f}}_{\rm{cut}}=0.45$ & $P^{\rm{f\bar
 f}}_{\rm{z,cut}}=700\rm{GeV}$ & $Y_{\rm C}=0.6$ & $Y_{\rm C}=1.2$
\\
\hline
 Significance(with $100\rm{fb}^{-1}$) & 18.3 & 18.4 & 12.1 & 17.1
\end{tabular}
\\ \hline
\end{tabular}

\end{table}

Generally speaking, the significance of the LR model is always
greater than that of the SSM in the benchmark parameters we choose.
$S_{RFB}$ and $S_{OFB}$ are not so sensitive to the
cuts as those of $S_C$ and $S_E$. Their maximum values are almost the same.
At the same time the optimized cuts
are the same for the two test models although their
magnitudes are different.  The reason is that the optimal cuts depend mainly
on the properties of the parton distribution function and mass of the ${Z^\prime}$.
Therefore the optimal cuts  are nearly independent on the chiral
properties of ${Z^\prime}$ coupling to fermions. The optimized cuts
 obtained from one specific ${Z^\prime}$ model are applicable to any other ${Z^\prime}$
model with the same ${Z^\prime}$ mass.

In LR model or SSM,
$A_{OFB}$ and $A_{RFB}$ can obtain almost the same highest
significance values. Significance of $A_E$ is smaller and
significance of $A_C$ is the smallest. The reason is that $A_E$ and
$A_C$ defined in the laboratory frame are diluted by the
longitudinal boosts from the partonic level to the hadron level. $A_C$
is even smaller because it includes more symmetric backgrounds than
that of $A_E$. Note that these results are based on the assumption that the
the final $f\bar{f}$ pair can be completely reconstructed. In the real
experiment, taking the top quark pair as the example, the momentum precision of the
top pair will be limited by the missing neutrino
when using the top semi-leptonically decaying mode \cite{Aaltonen:2011kc}. While $A_E$ and $A_C$ can be free from this problem because they utilize only the top or anti-top hadronic decay mode.

\subsection{Discriminating $Z^\prime$ models utilizing the optimized asymmetry}

Based on the optimal cuts obtained above, we can
discriminate different $Z^\prime$s via the precise asymmetry measurements at the LHC. In this part we calculated
the asymmetries and compare the optimal and un-optimal cases. The results for all
asymmetries can be deduced in the same procedure.
Note that the optimized cuts are almost the same for different ${Z^\prime}$ models provided that
the ${Z^\prime}$ mass is the same.

To identify the candidate ${Z^\prime}$ model,
the measured asymmetry should be compared with the theoretical predictions.
In our analysis, asymmetries by theoretical predictions with errors are presented as the
two-dimensional plot, similar to that in Ref. \cite{arXiv.0910.1334}.

\begin{figure}[htbp]
\begin{center}
\includegraphics[width=0.40\textwidth]
{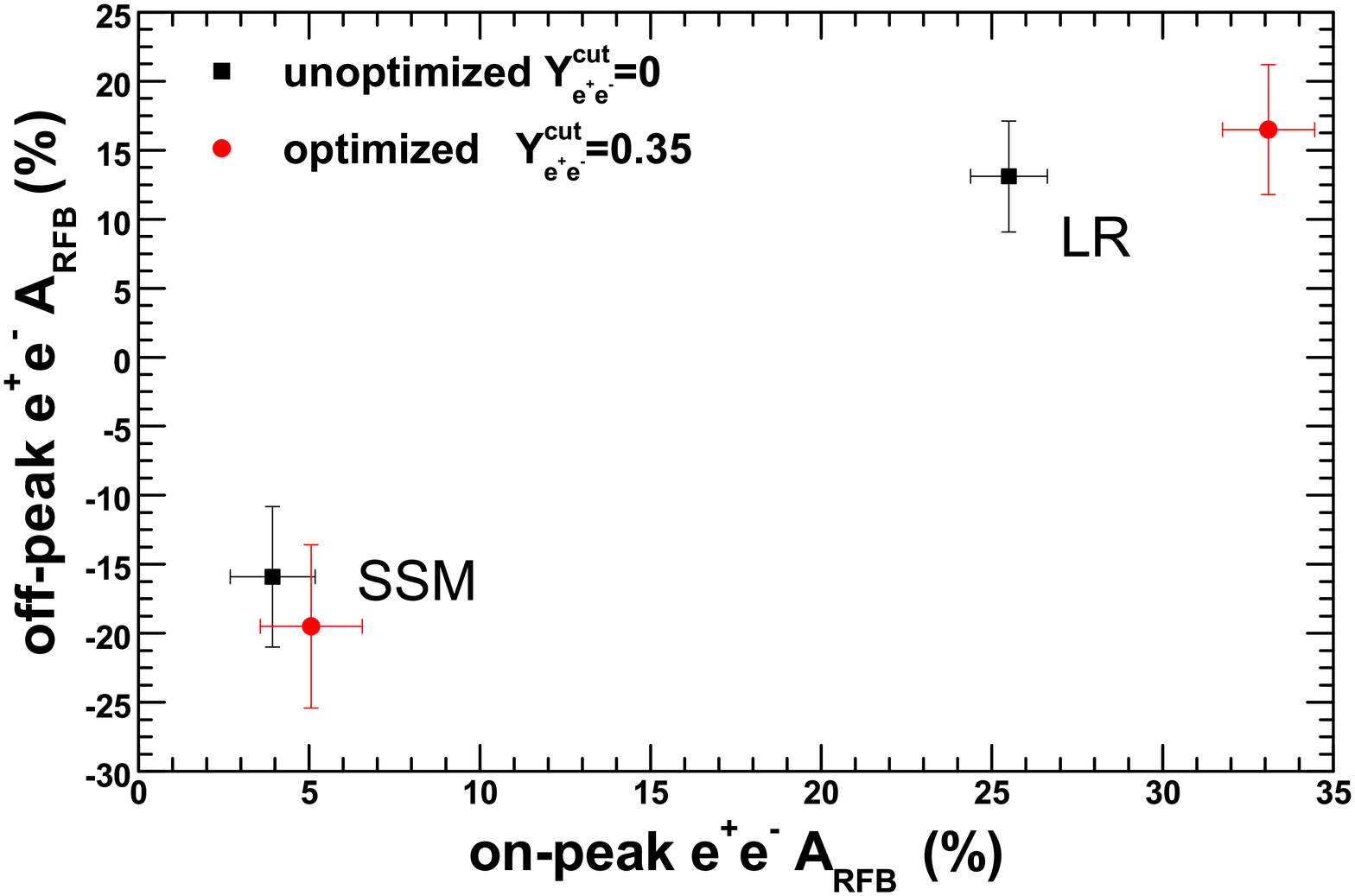}
\end{center}
\begin{center}
\includegraphics[width=0.40\textwidth]
{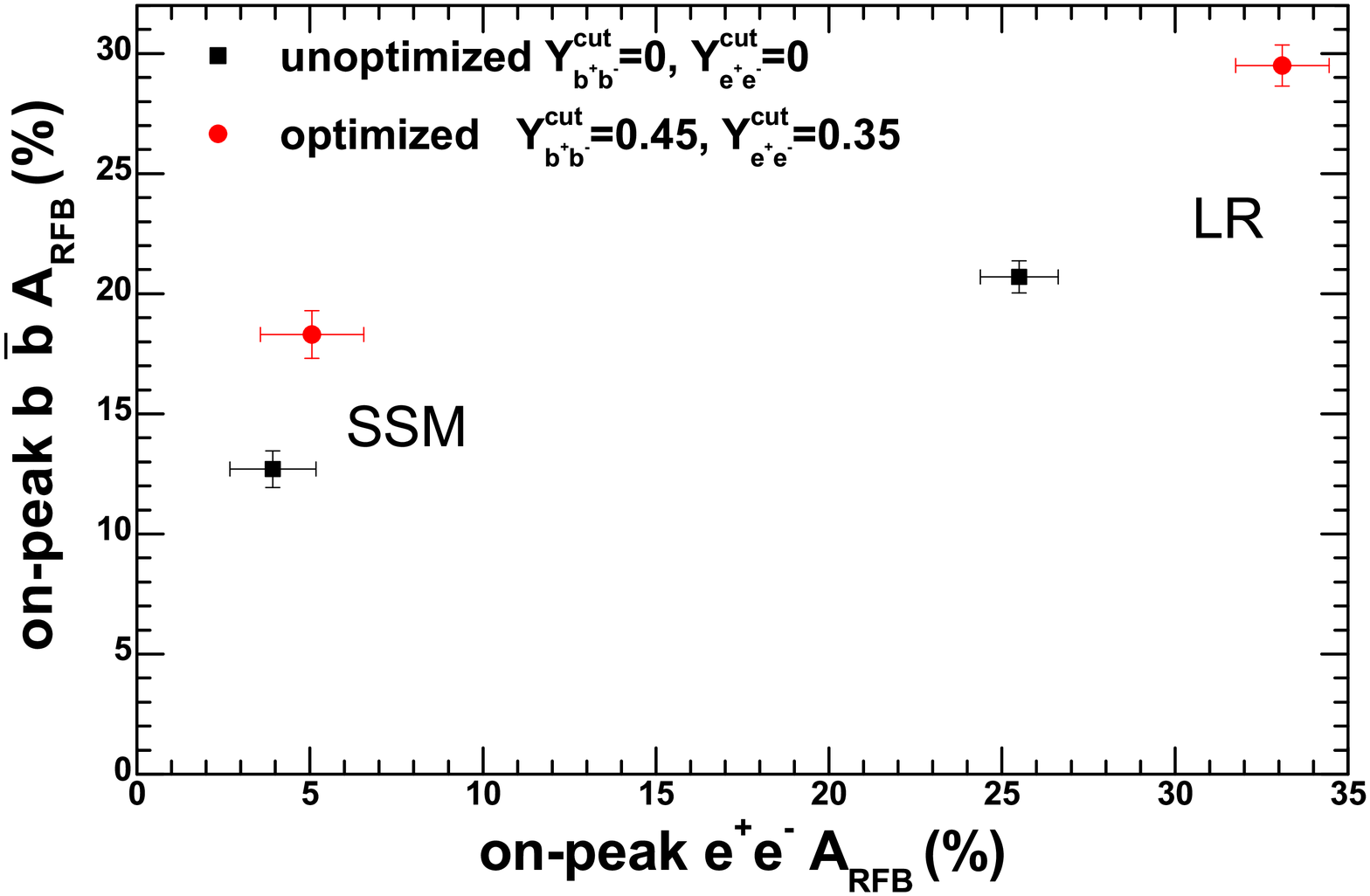}
\includegraphics[width=0.40\textwidth]
{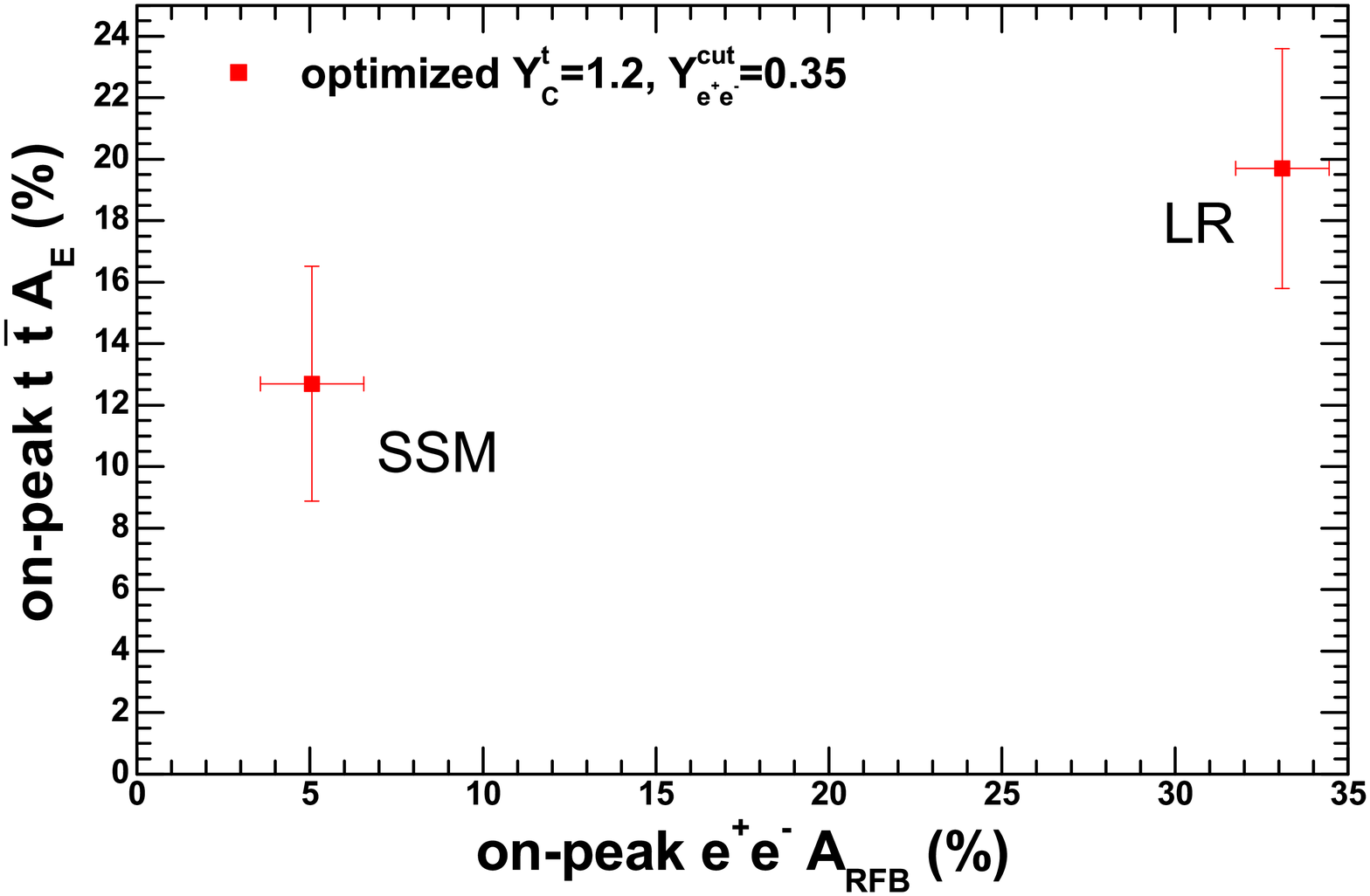}
\end{center}
\caption{\label{Dimention2Plots} Two dimensional plots of asymmetries $A_{RFB}$s  for off- and
on-peak $e^+e^-$ events, $A_{RFB}$s for on-peak $b\bar b$ and $e^+e^-$ events, and
$A_E$ and $A_{RFB}$ for on-peak $t\bar t$ and $e^+e^-$ events respectively.
Both optimized and un-optimized results for $A_{RFB}$ and optimized ones for
$A_E$ with
error bars in the LR model and SSM are presented. }
\end{figure}

In Fig. \ref{Dimention2Plots} we show the different asymmetries for both
LR model and the SSM. Central values are calculated in both optimized and
un-optimized cases. The error bar are estimated according to the formula
\begin{eqnarray}
{\rm{err}}\equiv \sqrt{\frac{4N^F
N^B}{N^3}}=\frac{1}{\sqrt{N}}\sqrt{1-(\frac{N^A}{N})^2}\cong \frac{1}{\sqrt{\mathcal
{L}\sigma\epsilon_{{f\bar f}}}}.
\end{eqnarray}
Here $N^F/N^B$ is the forward/backward events, $N^A=N^F-N^B$ is the
asymmetric events and $N$ is the total events. The relation between err and significance is
$S_{A}=A_{\rm{FB}}/\rm{err}$, where $S_A$ is the significance with reconstruction efficiency.
In our estimation, the $b\bar b$ and $t\bar t$ reconstruction
efficiencies are taken as $\epsilon_{b\bar b}=0.36$ and $\epsilon_{t\bar
t}=0.075$ respectively \cite{arXiv.0910.1334}.
For the $b\bar b$ and $t\bar t$ final
states, next-to-leading order QCD contribution to the asymmetric
cross sections are included. For the $b\bar b$ final state, QCD NLO
contribution to on-peak $A_{\rm{RFB}}$ is 1.38\% for the optimized
case and 0.96\% for the un-optimized $Y^{\rm{f\bar f}}_{\rm{cut}}=0$
case. The contribution is a little bit larger than the statistic error (see the
left-bottom diagram of Fig. \ref{Dimention2Plots}), so this effect should  be
taken into account. For the $t\bar t$ final states, QCD NLO
contribution to on-peak $A_{\rm{E}}$ is 0.53\% for the optimized
case, which is much less than the statistic error (see the
right-bottom diagram of Fig. \ref{Dimention2Plots}), so this effect can be
neglected.

From the Fig. \ref{Dimention2Plots}, the two models give the apparently different predictions for
two asymmetries. Even without optimal cuts, the asymmetries can be utilized to discriminate
models in this case. However in the real case, the asymmetry difference for various models
might be small. In this case the optimal cuts can help to improve the capacity to discriminate
models.
From the figure it is obviously that the central values of the asymmetries
separate more apart in both the LR model and
the SSM. However, due to the decreasing of the statistics, the error bar becomes a little bit
larger for the optimized case.

\section{Discussions and conclusions}\label{CONCLUSION}

In this paper we investigated how to utilize the asymmetry measurements at the LHC to discriminate
underlying dynamics, by taking $Z^\prime$ model as the example. Unlike LEP and Tevatron, the LHC is a symmetric
proton-proton collider, thus the asymmetry at the LHC has the unique feature which should be studied in detail.
In literature several asymmetries have been proposed
at the LHC, namely rapidity-dependent forward-backward asymmetry,
one-side forward-backward asymmetry, central charge asymmetry and
edge charge asymmetry, with $\ell^+\ell^-$, $b\bar b$ and $t\bar t$
as the final states.  Based on these works, we
stepped further on to analyze  how to optimize the asymmetries in the
left-right model and the sequential standard model.
In the calculations with $b\bar b$ and $t\bar t$ final states, the QCD-induced
higher order contributions to the asymmetric cross section were also included.
For each kind of final states, we estimated the four kinds of asymmetries and
especially the optimal cuts usually associated with the definition of the asymmetry.
Our studies showed that the optimal cut is stable
for different ${Z^\prime}$ model provided that the ${Z^\prime}$ mass is equal. The numerical results indicated
that the capacity to discriminate $Z^\prime$ models can be improved by imposing the optimal
cuts.

In this paper only $Z^\prime$ models of left-right
and sequential standard model were investigated as
the examples. However the optimization obtained from these two examples is
suitable for any kind of $Z^\prime$ models provided that they have the
same $Z^\prime$ mass. The
$Z^\prime$ mass throughout this paper is assumed to be 1.5 TeV as the benchmark
parameter. If the
$Z^\prime$ mass is other than 1.5 TeV, the optimal condition should
be investigated in the same procedure. Moreover precise asymmetry measurement at the LHC can
be utilized to scrutinize any new
dynamics beyond the standard model, not limited to the $Z^\prime$ case.

\section*{Acknowledgment}
This work was supported in part by the Natural Science Foundation
of China (No 11075003).

\end{document}